# Mechanism-resolved phase-field fracture of composite shells with a certified admissible constitutive operator

Altaf Ahmad Lone[a], R. T. Durai Prabhakaran[a], Timon Rabczuk[b,c], Xiaoying Zhuang[d]

[a] *Department of Mechanical Engineering, Indian Institute of Technology Jammu, Jagti, NH-44, Jammu, 181221, India*

[b] *Institute of Computational Mechanics and Artificial Intelligence, Fudan University, Shanghai,* 200433, China

[c] *Institute of Structural Mechanics, Bauhaus-Universität Weimar, Marienstraße 15, Weimar, 99423, Germany*

[d] *Chair of Computational Science and Simulation Technology, Institute of Photonics, Leibniz University Hannover, Appelstraße 11A, Hannover, 30167, Germany*

**Abstract**

Phase-field models of fracture in fibre-reinforced shells are usually formulated against a single closed-form stored energy, so that the stress, the consistent tangent and the plane-stress condensation are derived and verified for that expression alone. In the formulation developed here the constitutive law enters instead as a replaceable operator. The Green–Lagrange strain of a geometrically exact Reissner–Mindlin shell is pulled back through the Cholesky factor of the reference metric and resolved on four mutually orthogonal invariants with an exact closure identity, so that the separation of the stored energy into fibre, inter-fibre and interaction channels is a change of basis rather than a modelling assumption. On a curved midsurface the metric, its Cholesky factor and the fibre direction depend on the thickness coordinate through the shifter $\mathbf{I} - \zeta\mathbf{b}$, and the invariants inherit that dependence. Every subsequent operation — the tension–compression split, the degradation of each channel by its own phase field, the plane-stress condensation, the consistent tangent and the assembly — is expressed in the channel potentials and their derivatives alone. The residuals of both damage parameterisations are stated explicitly, and the fibre–transverse interaction energy is stored once and degraded by the product of the fields whose mechanisms it couples. Six conditions define the admissible operators; within them, the two linear invariants are proved convex in the deformation gradient directly, the geometric tangent is proved independent of the material tangent, and the condensation is proved regular wherever the residual stiffness is positive. A three-tier protocol certifies the algebra, the solver state and the admission of a state as training data. Because the mechanism channels are signed invariant-group contributions rather than energies of separable subsystems, shielding under transverse contraction appears as an inactive damage constraint and is reported through the pointwise positive part of each channel. Numerical studies on notched IM7/8552 shells verify the discretisation in the thin and curved regimes, and show that fibre orientation governs the damage envelope and the load capacity while curvature drives a through-thickness fracture asymmetry that a single midsurface field cannot represent.

**Highlights**

- Shell phase-field fracture driven by a replaceable, certified constitutive operator
- Orthogonal transverse-isotropy invariant basis with an exact closure identity
- Curved-shell chain certified by chart invariance and a flat-limit control
- Mechanism-resolved residuals stated explicitly; interaction energy stored once
- Curvature-induced through-thickness asymmetry resolved by two surface fields

Keywords: Phase-field fracture; Composite shells; Transverse isotropy; Constitutive operator; Variational consistency; Certification

## 1. Introduction

### 1.1 Failure mechanisms in fibre-reinforced shells

Fibre-reinforced shells do not fail in one way. Fibre rupture, inter-fibre cracking of the matrix and interfacial separation each carry their own strength, their own toughness and their own dependence on the angle between the fibre and the load, and the mechanism that limits a structure changes with that angle [1,2]. A model carrying a single scalar damage variable can reproduce a load–displacement curve, but it cannot report which mechanism consumed the energy.

The variational approach to fracture replaces explicit crack tracking by minimisation of a total energy in which the crack surface is regularised over a length scale [3–5], an elliptic approximation in the sense of Ambrosio and Tortorelli [6]. Its discrete form is a coupled displacement–damage problem [7,8] governed by three choices that are constitutive rather than numerical. First, the stored energy must be split so that compression does not drive damage [9]; the split fixes the shape of the strength surface and with it the nucleation response, which is why several inequivalent splits remain in use [10]. Second, irreversibility must be imposed pointwise rather than penalised, since a penalty parameter otherwise enters the solution it is meant to constrain [11]. Third, the functional is not jointly convex in the two fields, so the solution strategy, whether monolithic, staggered or quasi-Newton, selects the branch that is found [12–16], a sensitivity that implementations in general-purpose codes make plain [17]. Lone et al. [18] cross-validate a phase-field implementation against an extended finite element treatment of the same cracking problem for that reason.

### 1.2 Phase-field modelling of thin structures and composites

Thin structures add a difficulty of their own. The phase field is carried on the midsurface while the strain varies through the thickness, so the split and the degradation must be evaluated at several thickness stations and integrated, which prevents the usual separation of the energy into membrane and bending parts [19]. The same requirement appears with local maximum-entropy approximants [20], with solid-shell and enhanced-assumed-strain kinematics [21,22], with finite-strain plates and shells built on element subdivision [23], and in isogeometric Reissner–Mindlin shells, where the spectral split acts on the full strain tensor and plane stress must therefore be enforced numerically [24]. Two consequences bear on the present work. Plane-stress enforcement is a constitutive operation performed at every quadrature point, so its regularity is a property of the energy rather than of the element. And one midsurface field cannot represent a crack that reaches one face before the other, which is why fracture resolved along the thickness requires more than one damage variable [25].

Anisotropy enters a phase-field model by two distinct routes. It can be placed in the crack surface density, by adding a structural tensor to the gradient term so that the regularised crack itself acquires a preferred direction [26,27], or it can be placed in the stored energy, by splitting it into channels that correspond to physical mechanisms [28–30]. The two are not interchangeable: the first turns the crack, the second decides which mechanism is loaded. Multi-field formulations follow the second route and give each mechanism its own phase field and its own toughness, with the channel definitions frequently taken from an established failure criterion [31–33], while interfacial failure is treated by coupling to a cohesive zone [34,35] or by an additional interface field. Lone et al. [36] combine the composite phase-field description with elasto-plasticity and nonlocal regularisation, and the approach extends to viscoelastic–viscoplastic short-fibre composites under hygrothermal loading [37]. What these formulations share is that the channels and their closed-form expressions are fixed when the model is derived, so the stress, the tangent and the condensation are obtained once, for that expression.

### 1.3 Constitutive operators and data-driven material modelling

Constitutive modelling has moved in the opposite direction. Input-convex networks [38] make it possible to learn an energy that is polyconvex and anisotropic by construction [39–42]; stress labels can be avoided altogether by discovering the law from full-field kinematics and reaction forces [43]; and equilibrium itself can serve as the training loss, so that the residual structure of the finite element problem supervises the network [44–46]. A parallel line replaces the discretisation rather than the material law, minimising the potential energy over a network ansatz [47–50]; Lone et al. [51] follow that route in a multiscale setting, training a deep-energy network on finite element and experimental data to reproduce the thermo–viscoelastic–viscoplastic response of a unidirectional composite. All of this presumes that the surrounding boundary-value problem will accept whatever energy is supplied. A shell fracture solver written against one closed form does not: exchanging the energy means re-deriving the split, the stress, the consistent tangent and the plane-stress condensation, and re-verifying each of them. That obstacle is structural rather than computational, and it is what this work addresses.

## 1.4 Aim and scope

This work develops a phase-field fracture formulation for transversely isotropic composite shells in which the constitutive law enters as a replaceable operator. The Green–Lagrange strain of a geometrically exact Reissner–Mindlin shell is normalised by the Cholesky factor of the reference metric and expressed on four orthogonal invariants whose closure identity, Eq. (9), is exact. Because it is exact, the separation of the stored energy into a fibre channel, an inter-fibre channel and a fibre–transverse interaction channel is a change of basis, and the three channel potentials become the only material input. Everything downstream of them is written in those potentials and their derivatives, so the operator can receive either a closed form or a learned representation.

The contributions are as follows.

1. An orthogonal invariant basis for transverse isotropy is constructed in the shell setting, with an exact closure identity, so that the mechanism channels partition the strain norm without residue. The basis is carried through the thickness by the shifter of Eq. (2), so it is defined on curved as well as flat midsurfaces, and its chart invariance is certified in Section 7.4.
2. The coupled problem is derived from a single potential in which the fibre–transverse interaction energy is stored once and degraded by the product of the fields whose mechanisms it couples. The residuals of both damage parameterisations are stated explicitly in Eqs. (24) and (25), and the mechanism coupling appears in the residual as well as in the Hessian.
3. The formulation is representation-independent, and the two properties most at risk in an exchange of representation are proved to survive it: the geometric part of the tangent is nonzero and independent of the material tangent (Appendix A.6), and the plane-stress condensation stays regular wherever the residual stiffness is positive (Appendix A.5). Admissibility is stated as six conditions in Definition A.1. The two linear invariants are proved convex in the deformation gradient directly (Appendix A.3); the corresponding statement for the two quadratic invariants is left open and discharged by a pointwise certificate (Appendix A.4).
4. The shell discretisation of Section 5 is verified in the thin and curved regimes before any fracture result computed on it is interpreted, under a three-tier protocol that separates what is verified algebraically, what is verified at a solver state, and what qualifies a state for use as training data.
5. The formulation is exercised on notched composite shells, where fibre orientation and the interface field control distinct features of the fracture response, and where curvature drives a through-thickness asymmetry that a single midsurface field cannot represent.

The scope is bounded in one respect. The learned representation is admissible and certified, and its capacity to reproduce the closed form is measured in Appendix A.10, but it is not deployed: every result reported here is produced by the closed-form seat, with the network identically zero. Predictive learned constitutive modelling on this socket is the subject of a companion study.

Section 2 presents the shell kinematics, the strain measure and the through-thickness shifter. Section 3 develops the invariant basis, the mechanism decomposition, the two damage parameterisations, the stress and tangent, the plane-stress condensation and the crack driving force. Section 4 derives the coupled Euler–Lagrange equations and the consistent tangent from a single potential, and Section 5 gives the discretisation and the solution scheme. Section 6 states the three-tier certification protocol, Section 7 presents the numerical studies and Section 8 summarises the findings. Appendix A collects the admissibility definition and the proofs on which Sections 3–6 rely.

## 1.5 Notation

Direct notation is used throughout. Light-face italic characters denote scalars and scalar-valued functions, for example the shell thickness $h$, the reaction force $F$ and the stored energy $\Psi$. Vectors and second-order tensors are written in italic boldface, for example $\mathbf{m}$, $\mathbf{E}$ and the deformation gradient $\mathbf{F}$; fourth-order tensors and linear operators between tensor spaces are written in blackboard bold, for example $\mathbb{C}$. The single and double contractions of two second-order tensors are written $\mathbf{AB}$ and $\mathbf{A}:\mathbf{B}$, the dyadic product as $\mathbf{a} \otimes \mathbf{b}$, and the transpose as $(\cdot)^{\mathsf{T}}$. A hat denotes a quantity referred to the metric-normalised configuration of Eq. (6), as in $\hat{\mathbf{E}}$, and a superscript $ps$ denotes a plane-stress condensed quantity, as in $\mathbb{C}^{ps}$. The operators $\mathrm{tr}(\cdot)$, $\mathrm{dev}(\cdot)$ and $\det(\cdot)$ denote the trace, the deviatoric part and the determinant; $\|\cdot\|$ denotes the Euclidean norm of a vector and the Frobenius norm of a tensor, and $\succcurlyeq \mathbf{0}$ positive semi-definiteness. Macaulay brackets are $\langle x\rangle_+ = (x+|x|)/2$ and $\langle x\rangle_- = (x-|x|)/2$.

Greek indices range over {1,2} and Latin indices over {1,2,3}, with summation implied over repeated indices unless stated otherwise. Two symbols carry a second, declared meaning. The subscript $\alpha$ on the mechanism quantities $\Psi_\alpha$, $H_\alpha$, $G_c^\alpha$, $\ell_\alpha$ and $\mathcal{I}_\alpha$ labels the three channels, $\alpha \in \{f, mx, int\}$ for fibre, inter-fibre and interaction, and is never summed; elsewhere $\alpha$ is a surface coordinate index. The symbol $\beta$ standing alone denotes the interfacial phase field and is never used as an index. The operator $\delta(\cdot)$ denotes the first variation, while $\delta$ standing alone denotes the prescribed edge displacement of the benchmarks of Section 7.

## 2. Geometrically exact shell kinematics

Consider a shell occupying the reference configuration $\Omega_0 = \mathcal{S}_0 \times [-h/2,\, h/2]$, where $\mathcal{S}_0 \subset \mathbb{R}^3$ is the reference midsurface and $h$ is the shell thickness, parameterised by curvilinear coordinates $\boldsymbol{\xi} = (\xi^1, \xi^2)$ with position vector $\mathbf{r}_0$.

The covariant tangent basis $\mathbf{A}_\alpha = \partial \mathbf{r}_0 / \partial \xi^\alpha$ defines the reference midsurface metric $G_{\alpha\beta} = \mathbf{A}_\alpha \cdot \mathbf{A}_\beta$ with inverse $G^{\alpha\beta}$. The unit surface normal is $\mathbf{N} = (\mathbf{A}_1 \times \mathbf{A}_2)/\| \mathbf{A}_1 \times \mathbf{A}_2 \|$, and the curvature is described by the second fundamental form $b_{\alpha\beta} = \mathbf{N} \cdot \mathbf{A}_{\alpha,\beta}$. These satisfy the Gauss–Weingarten relations

$$\mathbf{A}_{\alpha,\beta} = \Gamma^\gamma_{\alpha\beta} \mathbf{A}_\gamma + b_{\alpha\beta} \mathbf{N}, \qquad \mathbf{N}_{,\alpha} = -b_{\alpha\beta} G^{\beta\gamma} \mathbf{A}_\gamma \tag{1}$$

where $\Gamma^\gamma_{\alpha\beta}$ are the Christoffel symbols of the surface metric.

The shell basis at a thickness station $\zeta$ follows from the Weingarten relation of Eq. (1) without approximation,

$$\mathbf{A}_\alpha(\zeta) = \left(\delta^\gamma_\alpha - \zeta\, b^\gamma_\alpha\right)\mathbf{A}_\gamma, \qquad \mathbf{Z}(\zeta) = \mathbf{I} - \zeta \mathbf{b}, \qquad J(\zeta) = \det \mathbf{Z}(\zeta) = 1 - 2H\zeta + K\zeta^2 \tag{2}$$

with $H = 1/2\,\mathrm{tr}\mathbf{b}$ and $K = \det\mathbf{b}$ the mean and Gaussian curvatures. The scalar $J(\zeta)$ is the through-thickness Jacobian that appears in the volume measure of Eqs. (19) and (33). The reference metric of Eq. (5) is therefore itself a function of the thickness coordinate, $\mathbf{G}_0(\zeta) = \mathrm{diag}(\mathbf{Z}^\mathsf{T}(\zeta)\,\mathbf{G}\,\mathbf{Z}(\zeta),\, 1)$ with $\mathbf{G} = \left[G_{\alpha\beta}\right]$, and its Cholesky factor $\mathbf{L}(\zeta)$, the pull-back of Eq. (6) and the normalised fibre direction $\widehat{\mathbf{m}}(\zeta)$ are evaluated at every station rather than once at the midsurface. The four invariants of Eq. (8) consequently vary through the thickness whenever $b_{\alpha\beta} \neq 0$, which is the route by which curvature reaches the constitutive law. For $b_{\alpha\beta} = 0$ the shifter is the identity, every station collapses onto the midsurface expression and the flat formulation is recovered exactly.

The fibre field is constructed on the same basis. On a curved midsurface a fixed ambient direction projected into each tangent plane does not define a ply of constant orientation, because the projection angle varies over the surface. The fibre direction is therefore built intrinsically: the axis of the surface is projected into the tangent plane and rotated by the ply angle $\theta$ about the normal, which produces a helix on a cylinder and a loxodrome on a sphere from one expression, and is normalised at every station so that $\mathbf{m}^\mathsf{T}\mathbf{G}_0(\zeta)\,\mathbf{m} = 1$, equivalently $\| \widehat{\mathbf{m}} \| = 1$ in Eq. (6). The two conventions are not equivalent: over a 50 mm patch at $R = 100$ mm the projected convention departs from a constant-angle winding by up to 70.6°, so an orientation-dependent result computed with it would not describe the ply layout it is attributed to.

### 2.1 Kinematics and strain measure

The current configuration is reconstructed from the midsurface displacement $\mathbf{u}$ and an independent director field $\mathbf{d}$, parameterised by two rotational degrees of freedom $\theta^\alpha$,

$$\mathbf{x}(\xi^\alpha, \zeta) = \mathbf{r}_0 + \mathbf{u} + \zeta \mathbf{d}, \qquad \mathbf{d} = \mathbf{N} + \theta^\alpha \mathbf{A}_\alpha \tag{3}$$

valid for $-h/2 \le \zeta \le h/2$. Equation (3) is the geometrically nonlinear Reissner–Mindlin kinematic assumption used throughout; the director is an extensible vectorial update rather than an exact finite rotation, with direct consequences for the structural tangent (Appendix A.6). Differentiating gives the deformed shell basis $\mathbf{a}_\alpha = \mathbf{A}_\alpha + \mathbf{u}_{,\alpha} + \zeta \mathbf{d}_{,\alpha}$ and $\mathbf{a}_3 = \mathbf{d}$, so that the shell deformation gradient and its reference counterpart are

$$\mathbf{F} = [\,\mathbf{a}_1,\ \mathbf{a}_2,\ \mathbf{d}\,], \qquad \mathbf{F}_0 = \left[\,\mathbf{A}_1 + \zeta \mathbf{N}_{,1},\ \mathbf{A}_2 + \zeta \mathbf{N}_{,2},\ \mathbf{N}\,\right] \tag{4}$$

with reference metric and Green–Lagrange strain

$$\mathbf{G}_0 = \mathbf{F}_0^{\mathsf{T}}\mathbf{F}_0, \qquad \mathbf{E} = 1/2\,(\mathbf{F}^{\mathsf{T}}\mathbf{F} - \mathbf{G}_0) \qquad (5)$$

The columns of $\mathbf{F}_0$ are the shifted basis vectors of Eq. (2), so $\mathbf{G}_0$ is the thickness-dependent metric introduced there. Under a superposed rigid-body rotation $\mathbf{F} \mapsto \mathbf{QF}$, $\mathbf{Q} \in SO(3)$, one has $(\mathbf{QF})^{\mathsf{T}}(\mathbf{QF}) = \mathbf{F}^{\mathsf{T}}\mathbf{F}$, so $\mathbf{E}$ is unchanged and Eq. (5) is objective.

## 3. Constitutive model

### 3.1 Orthogonal invariant basis

The Green–Lagrange strain of Eq. (5) is not the natural constitutive variable of a transversely isotropic lamina, because it is referred to a metric that is neither Euclidean nor orthogonal. Writing the Cholesky factorisation of the reference metric as $\mathbf{G}_0 = \mathbf{L}\mathbf{L}^{\mathsf{T}}$, the normalised strain and fibre direction are

$$\hat{\mathbf{E}} = \mathbf{L}^{-1}\mathbf{E}\mathbf{L}^{-\mathsf{T}}, \qquad \hat{\mathbf{m}} = \mathbf{L}^{\mathsf{T}}\mathbf{m}, \qquad \hat{\mathbf{m}} \cdot \hat{\mathbf{m}} = 1 \qquad (6)$$

with $\mathbf{m}$ normalised in the metric $\mathbf{G}_0$, so that the constitutive problem is posed entirely in Euclidean space. The fibre and transverse projectors,

$$\mathbf{P}_f = \hat{\mathbf{m}} \otimes \hat{\mathbf{m}}, \qquad \mathbf{P}_t = \mathbf{I} - \mathbf{P}_f \qquad (7)$$

satisfy $\mathbf{P}_f^2 = \mathbf{P}_f$, $\mathbf{P}_f + \mathbf{P}_t = \mathbf{I}$ and $\mathbf{P}_f\mathbf{P}_t = \mathbf{0}$, and induce the orthogonal decomposition $\mathbb{R}^3 = \mathrm{span}(\hat{\mathbf{m}}) \oplus \mathrm{span}(\hat{\mathbf{m}})^{\perp}$. Four mutually orthogonal constitutive coordinates follow, namely the fibre extension $I_f$ and the transverse invariants $I_t, J_t$ and $J_{ft}$,

$$\begin{aligned} I_f &= \hat{\mathbf{m}}^{\mathsf{T}}\hat{\mathbf{E}}\,\hat{\mathbf{m}}, & \mathbf{E}_t &= \mathbf{P}_t\hat{\mathbf{E}}\mathbf{P}_t, & I_t &= \mathrm{tr}\mathbf{E}_t, \\ \mathrm{dev}\mathbf{E}_t &= \mathbf{E}_t - 1/2\,I_t\mathbf{P}_t, & J_t &= \mathrm{dev}\mathbf{E}_t : \mathrm{dev}\mathbf{E}_t, & J_{ft} &= 2\big(\mathbf{P}_f\hat{\mathbf{E}}\mathbf{P}_t\big) : \big(\mathbf{P}_f\hat{\mathbf{E}}\mathbf{P}_t\big) \end{aligned} \qquad (8)$$

where dev acts within the two-dimensional transverse subspace. The set $\mathcal{I} = \{I_f, I_t, J_t, J_{ft}\}$ satisfies the exact orthogonal closure identity

$$\hat{\mathbf{E}} : \hat{\mathbf{E}} = I_f^2 + 1/2\,I_t^2 + J_t + J_{ft} \qquad (9)$$

which follows by expansion in the eigenbasis of $\mathbf{P}_f$. Equation (9) is the transversely isotropic analogue of the Spencer–Boehler structural-tensor invariant basis used in anisotropic phase-field fracture [26,37].

### 3.2 Mechanism decomposition and the constitutive operator

Because Eq. (9) closes exactly, the Helmholtz free energy may be decomposed on the invariant groups without residue,

$$\Psi = \Psi_f\big(I_f\big) + \Psi_{mx}(I_t, J_t) + \Psi_{int}\big(J_{ft}\big) \qquad (10)$$

the fibre mechanism responding to fibre extension alone, the inter-fibre mechanism to transverse dilatation and distortion, and the interaction mechanism to fibre–transverse coupling.

Nothing in Eq. (10) requires $\Psi_f$, $\Psi_{mx}$ and $\Psi_{int}$ to be fixed closed-form expressions. Each is a map $\Psi_\alpha : \mathcal{I}_\alpha \to \mathbb{R}$, $\alpha \in \{f, mx, int\}$, where $\mathcal{I}_\alpha \subseteq \mathcal{I}$ is the invariant group of that channel. Equations (11)–(35) below — the tension–compression split, the degradation, the plane-stress condensation, the residual, the consistent tangent $\mathbf{K} = D^2\Pi$ and the finite element assembly — are stated purely in terms of $\Psi_\alpha$ and its derivatives, and hold unchanged for any $\Psi_\alpha$ satisfying the admissibility conditions of Definition A.1. An analytic $\Psi_\alpha$ may therefore be replaced by a learned one without altering the shell mechanics. One route to such an operator that requires no direct stress measurement is the equilibrium-based training paradigm of Li and Chen [44], extended to elastoplastic history dependence by Li et al. [45] and to three-dimensional anisotropic hyperelasticity with data-driven symmetry discovery by Li et al. [46]; Appendix A.9 makes the connection precise for the present shell model.

The operator deployed in Section 7 is a closed form that is quadratic in each channel, with the invariant coefficients listed in Fig. 2(c) and the two coercivity barriers of Fig. 3(c). Its inter-fibre channel carries one term whose sign is not fixed, namely a star-convex contribution in $I_t$ that is active only for $I_t < 0$, in the sense of the star-convex decomposition of Vicentini et al. [10]. It makes the transverse compressive strength settable independently of the tensile one by removing driving force under transverse contraction, so that $\Psi_f$, $\Psi_{mx}$ and $\Psi_{int}$ are contributions of invariant groups rather than energies of separable subsystems: their sum is the stored energy, but individually they are signed, and they are not unique, since any invariant basis for transverse isotropy admits reparameterisations that move content between groups while leaving the total unchanged [52,53]. At a fibre angle of $90°$ the inter-fibre contribution is negative at $55{,}077$ of $121{,}074$ quadrature points, against a decomposition that closes to $1.3 \times 10^{-16}$ of the total. The negative values are therefore a property of the star-convex term and not a defect of the split, and no coercivity barrier is involved, since the barriers lie outside the mechanism channels.

### 3.3 Tension–compression split and damage parameterisation

Fracture evolves only in tension, so each mechanism is split into fracture-active and fracture-inactive parts,

$$\Psi_\alpha = \Psi_\alpha^+ + \Psi_\alpha^-, \qquad \Psi^+ = \sum_\alpha \Psi_\alpha^+, \qquad \Psi^- = \sum_\alpha \Psi_\alpha^- \tag{11}$$

with a Macaulay-bracket split on the two linear invariants $I_f$ and $I_t$. The quadratic invariants $J_t$ and $J_{ft}$ of Eq. (8) are sums of squares and therefore nonnegative, so no sign split applies to them and they are taken as fully fracture-active, $\Psi_\alpha^- = 0$ for the channels they drive.

The formulation carries two bulk damage slots $d_1$ and $d_2$ together with the interfacial field $\beta$ of Eq. (21), and the assignment of those slots to physical quantities is a modelling choice that the variational structure does not constrain. Two assignments are used.

In the **surface-resolved** parameterisation, $(d_1, d_2) = (d_u, d_l)$ are independent phase fields on the upper and lower shell surfaces, which combine through the thickness as

$$d(\zeta) = c_u(\zeta)\, d_u + c_l(\zeta)\, d_l, \qquad c_u + c_l = 1, \qquad c_u(h/2) = 1, \qquad c_u(-h/2) = 0 \tag{12}$$

with the linear interpolant $c_u = 1/2 + \zeta/h$ used throughout. The fracture-active energy is degraded pointwise by a single field,

$$g(d) = (1-d)^2 + k, \qquad \Psi_d = g(d)\, \Psi^+ + \Psi^- \tag{13}$$

with residual stiffness $0 < k \ll 1$, so that $g(0) = 1$, $g(1) = k$ and $g' \le 0$. Equation (13) is the AT2 degradation of the Ambrosio–Tortorelli regularisation.

In the **mechanism-resolved** parameterisation, $(d_1, d_2) = (\varphi_f, \varphi_{if})$ and each slot degrades its own channel, while the interaction channel is degraded by the product of the three fields whose mechanisms it couples,

$$\Psi_d = g(\varphi_f)\, \Psi_f^+ + g(\varphi_{if})\, \Psi_{mx}^+ + g(\varphi_f)\, g(\varphi_{if})\, g_\beta(\beta)\, \Psi_{int}^+ + \Psi^- \tag{14}$$

The two parameterisations are not interchangeable, and each carries its own residuals, stated in Section 4 as Eq. (24) and Eq. (25). They differ in more than labelling. The surface-resolved slots enter through the thickness interpolants $c_u$ and $c_l$ and therefore resolve variation along $\zeta$, whereas the mechanism-resolved slots are uniform through the thickness and instead resolve which mechanism dissipates. The interaction channel of Eq. (14) is degraded by a product of fields, which the single-degradation form of Eq. (13) cannot express, and under Eq. (14) each field carries its own toughness $G_c^\alpha$ and its own regularisation length $\ell_\alpha$, so the mechanisms compete on both counts. Sections 7.11–7.13 use the mechanism-resolved parameterisation and Section 7.14 the surface-resolved one.

**Remark 1.** The product form acting on the interaction channel in Eq. (14) is required by the content of that channel. The potential $\Psi_{int}$ is built on $J_{ft}$, Eq. (8), which measures fibre–transverse coupling and therefore cannot survive the loss of either constituent. Degrading it by a single field would leave the coupling active after one of the two had already failed.

### 3.4 Stress, consistent tangent and plane-stress condensation

The second Piola–Kirchhoff stress and the material tangent follow from the degraded density by differentiation,

$$\mathbf{S} = \frac{\partial \Psi_d}{\partial \mathbf{E}}, \qquad \mathbb{C} = \frac{\partial \mathbf{S}}{\partial \mathbf{E}} = \frac{\partial^2 \Psi_d}{\partial \mathbf{E}\, \partial \mathbf{E}} \qquad (15)$$

A plane-stress response is enforced constitutively at every quadrature point by solving $S_{33} = 0$ locally for the thickness strain and condensing on the 33-component,

$$S_{33}\big(E_{33}^*; E_{\alpha\beta}\big) = 0, \qquad S_{\alpha\beta}^{ps} = S_{\alpha\beta}|_{E_{33}=E_{33}^*}, \qquad \mathbb{C}_{\alpha\beta\gamma\delta}^{ps} = \mathbb{C}_{\alpha\beta\gamma\delta} - \frac{\mathbb{C}_{\alpha\beta 33}\, \mathbb{C}_{33\gamma\delta}}{\mathbb{C}_{3333}} \qquad (16)$$

where every entry is evaluated at the converged $E_{33}^*$ and the condensed operators $(\mathbf{S}^{ps}, \mathbb{C}^{ps})$ are those used in the finite element formulation. Equation (16) is the implicit-function-theorem condensation, and its regularity for any admissible $\Psi_\alpha$ rests on $\mathbb{C}_{3333} \neq 0$, proved in Appendix A.5.

### 3.5 Crack driving force

A channel contribution that is negative under transverse contraction would, if substituted directly into a damage residual, release damage there and allow that residual to change sign within a single element. The quantity used for reporting is therefore the pointwise positive part of the channel contribution,

$$H_\alpha = \langle \Psi_\alpha \rangle_+, \qquad \alpha \in \{f, mx, int\} \qquad (17)$$

evaluated at each quadrature point before any integration, so that the positive part is taken of the field and not of its integral. It is nonnegative by construction and coincides with $\Psi_\alpha$ wherever that channel is positive.

The damage residuals of Section 4 are driven by the fracture-active channel energies $\Psi_\alpha^+$ of Eq. (11) and not by $H_\alpha$. Where a channel is shielded, the corresponding residual is one-signed over the whole feasible direction and the field remains at its irreversibility bound, so shielding is realised as an inactive constraint in the Karush–Kuhn–Tucker system of Eq. (35) rather than as a modification of the driving term. The quantity $H_\alpha$ is the post-processed diagnostic through which it is reported.

**Remark 2.** Taking the positive part pointwise means that the channel driving forces no longer sum exactly to the degradable part of the stored energy once shielding is active. The cost is measured rather than assumed: the relative additivity gap is zero at fibre angles of 0°, 30° and 60°, and $2.0 \times 10^{-2}$ at 90°, where shielding suppresses the inter-fibre driving force over 45.5% of the quadrature points. It is reported together with every result that depends on it.

## 4. Variational formulation and linearisation

The governing equations follow from stationarity of a single total potential energy, which combines the degraded mechanical energy, the fracture regularisation of the two bulk damage fields, the crack surface energy of the interfacial field and the external potential,

$$\Pi(\mathbf{w}) = \Pi_{mech} + \Pi_{d_1} + \Pi_{d_2} + \Pi_\beta - W_{ext}, \qquad \mathbf{w} = (\mathbf{u}, \boldsymbol{\theta}, d_1, d_2, \beta) \qquad (18)$$

with

$$\Pi_{mech} = \int_{\mathcal{S}_0} \int_{-h/2}^{h/2} \Psi_d\,(\mathbf{E},\, d_1,\, d_2,\, \beta)\, J(\zeta)\, \mathrm{d}\zeta\, \mathrm{d}A \qquad (19)$$

where $J(\zeta)$ is the shifter of Eq. (2) and $\Psi_d$ is the degraded density of Eq. (13) or Eq. (14) according to the parameterisation in force. Equation (19) is the only place in the functional where stored energy appears, and its measure is abbreviated below as $\mathrm{d}V = J(\zeta)\, \mathrm{d}\zeta\, \mathrm{d}A$ on $\Omega_0 = \mathcal{S}_0 \times [-h/2, h/2]$. The bulk regularisation terms and the external work are

$$\Pi_{d_i} = \int_{\mathcal{S}_0} G_{c,i} \left( \frac{d_i^2}{2\ell_i} + \frac{\ell_i}{2} \parallel \nabla d_i \parallel^2 \right) \mathrm{d}A, \qquad i = 1,2 \qquad (20)$$

$$\Pi_\beta = \int_{\mathcal{S}_0} G_c^{int}\left(\frac{\beta^2}{2\ell_\beta} + \frac{\ell_\beta}{2}\ \| \nabla\beta \|^2\right)\mathrm{d}A, \qquad g_\beta(\beta) = (1-\beta)^2 + k_\beta \tag{21}$$

$$W_{ext} = \int_{\mathcal{S}_0} \mathbf{b}\cdot\mathbf{u}\,\mathrm{d}A + \int_{\Gamma_t} \bar{\mathbf{t}}\cdot\mathbf{u}\,\mathrm{d}\Gamma \tag{22}$$

with $0 < k_\beta \ll 1$. Under the surface-resolved parameterisation the two bulk fields are faces of the same crack and share the material values, $(G_{c,1}, \ell_1) = (G_{c,2}, \ell_2) = (G_c, \ell_c)$; under the mechanism-resolved parameterisation they carry the channel values $(G_{c,1}, \ell_1) = (G_c^f, \ell_f)$ and $(G_{c,2}, \ell_2) = (G_c, \ell_c)$. Because Eq. (14) already carries the interfacial degradation, $\Pi_\beta$ reduces to the crack surface energy of $\beta$ alone.

**Remark 3.** Degrading $\Psi_{int}^+$ by $g_\beta$ inside a separate interfacial potential while it also enters $\Psi_d$ would store the same driving energy twice, let two dissipation paths draw on it independently, and assign the two contributions different measures, one through the thickness and one over the midsurface. In the product form of Eq. (14) the interaction energy appears once and $\beta$ reduces it exactly as $\varphi_f$ and $\varphi_{if}$ do; the channel decomposition of $\Pi$ closes to $3\times10^{-16}$ of $\Pi$ at every load step of every run reported in Section 7.

Stationarity, $\delta\Pi = 0$ for all admissible $\delta\mathbf{w}$, gives five coupled Euler–Lagrange equations. Mechanical equilibrium reads

$$R_{mech}(\delta\mathbf{u}, \delta\boldsymbol{\theta}) = \int_{\Omega_0} \mathbf{S} : \delta\mathbf{E}\,\mathrm{d}V - \delta W_{ext} \tag{23}$$

with $\mathbf{S} = \partial\Psi_d/\,\partial\mathbf{E}$ evaluated at the degraded density in force. The bulk damage residuals depend on the parameterisation. Under the surface-resolved assignment, Eqs. (12), (13) and (20) give

$$\begin{aligned} R_{d_u}(\delta d_u) \quad &= \int_{\Omega_0} g'(d)\,c_u\,\Psi^+\,\delta d_u\,\mathrm{d}V + \int_{\mathcal{S}_0} G_c\left(\frac{d_u}{\ell_c}\,\delta d_u + \ell_c\,\nabla d_u\cdot\nabla\delta d_u\right)\mathrm{d}A, \\ R_{d_l}(\delta d_l) \quad &= \int_{\Omega_0} g'(d)\,c_l\,\Psi^+\,\delta d_l\,\mathrm{d}V + \int_{\mathcal{S}_0} G_c\left(\frac{d_l}{\ell_c}\,\delta d_l + \ell_c\,\nabla d_l\cdot\nabla\delta d_l\right)\mathrm{d}A \end{aligned} \tag{24}$$

and $\Psi_{int}^+$ carries no $\beta$-dependence in Eq. (13), so $R_\beta$ has no mechanical driving term and $\beta$ remains at its irreversibility bound. Under the mechanism-resolved assignment the residuals follow by differentiating Eq. (14) with respect to each field in turn,

$$\begin{aligned} R_{\varphi_f}(\delta\varphi_f) \quad &= \int_{\Omega_0} g'(\varphi_f)\left[\Psi_f^+ + g(\varphi_{if})\,g_\beta(\beta)\,\Psi_{int}^+\right]\delta\varphi_f\,\mathrm{d}V + \int_{\mathcal{S}_0} G_c^f\left(\frac{\varphi_f}{\ell_f}\,\delta\varphi_f + \ell_f\,\nabla\varphi_f\cdot\nabla\delta\varphi_f\right)\mathrm{d}A, \\ R_{\varphi_{if}}(\delta\varphi_{if}) \quad &= \int_{\Omega_0} g'(\varphi_{if})\left[\Psi_{mx}^+ + g(\varphi_f)\,g_\beta(\beta)\,\Psi_{int}^+\right]\delta\varphi_{if}\,\mathrm{d}V + \int_{\mathcal{S}_0} G_c\left(\frac{\varphi_{if}}{\ell_c}\,\delta\varphi_{if} + \ell_c\,\nabla\varphi_{if}\cdot\nabla\delta\varphi_{if}\right)\mathrm{d}A, \\ R_\beta(\delta\beta) \quad &= \int_{\Omega_0} g_\beta'(\beta)\,g(\varphi_f)\,g(\varphi_{if})\,\Psi_{int}^+\,\delta\beta\,\mathrm{d}V + \int_{\mathcal{S}_0} G_c^{int}\left(\frac{\beta}{\ell_\beta}\,\delta\beta + \ell_\beta\,\nabla\beta\cdot\nabla\delta\beta\right)\mathrm{d}A \end{aligned} \tag{25}$$

Two features of Eq. (25) are absent from Eq. (24). Each bulk field is driven by its own channel and, through the product, by the interaction channel weighted by the current state of the other two fields, so the mechanisms are coupled already in the residual and not only in the tangent. The interfacial driving term carries the factor $g(\varphi_f)\,g(\varphi_{if})$, so once either bulk mechanism has failed the interaction energy that $\beta$ was degrading has been removed and $\beta$ ceases to be driven, which is the discrete statement of Remark 1. In both parameterisations the driving terms carry the volume measure $\mathrm{d}V$ and the regularisation terms the surface measure $\mathrm{d}A$; the two are not interchangeable, and on a curved shell they differ at order $h/R$. Collecting all contributions gives $\mathbf{R}(\mathbf{w}) = \mathbf{0}$ with

$$\mathbf{R} = \begin{bmatrix} \mathbf{R}_\mathbf{u} \\ \mathbf{R}_\boldsymbol{\theta} \\ \mathbf{R}_{d_1} \\ \mathbf{R}_{d_2} \\ \mathbf{R}_\beta \end{bmatrix} = \mathbf{0} \tag{26}$$

whose blocks are the nodal coefficients of the linear forms above under the interpolation of Eq. (32).

## 4.1 Consistent linearisation

Equation (26) is solved by Newton's method. Since the residual is the first variation of a scalar potential, the tangent is the second variation of the same potential,

$$\mathbf{K}(\mathbf{w}) = D\mathbf{R}(\mathbf{w}) = D^2\Pi(\mathbf{w}) \qquad (27)$$

so that every block of the Jacobian originates from $\Pi$, in the manner of the stress-resultant shell tangent of Simo and Fox [54]. Linearising Eqs. (23) and (13) gives the mechanical tangent and the damage-linearisation term,

$$DR_{mech} = \int_{\Omega_0} (\delta\mathbf{E}:\mathbb{C}:\Delta\mathbf{E} + \mathbf{S}:D(\delta\mathbf{E})[\Delta\mathbf{w}])\,\mathrm{d}V, \qquad \mathbf{K}_{mech} = \mathbf{K}_{mat} + \mathbf{K}_{geo} \qquad (28)$$

$$D\left(\frac{\partial\Psi_d}{\partial d}\right) = g''(d)\,\Psi^+\,\Delta d + g'(d)\,D\Psi^+[\Delta\mathbf{E}] \qquad (29)$$

written here for the surface-resolved density of Eq. (13); the mechanism-resolved case follows from Eq. (14) in the same way. The first term of Eq. (29) gives the damage self-stiffness and, through $d(\zeta)$ in Eq. (12), the $d_u$–$d_l$ cross-term, and the second the mechanical–damage coupling. The geometric contribution $\mathbf{K}_{geo}$ depends on the current stress $\mathbf{S}$ and on the purely kinematic operator built from Eqs. (3) and (4), never on $\mathbb{C}$, and is nonzero for this kinematics regardless of how $\Psi_\alpha$ is represented (Appendix A.6). Because $\Psi_{int}^+$ depends on $\mathbf{E}$ but not on the bulk fields, linearising Eq. (21) couples $\beta$ to $\mathbf{u}$ and $\boldsymbol{\theta}$ through $\mathbf{E}$ rather than to $d_1$ and $d_2$. Collecting all contributions,

$$\mathbf{K} = \begin{bmatrix} \mathbf{K}_{\mathbf{uu}} & \mathbf{K}_{\mathbf{u}\boldsymbol{\theta}} & \mathbf{K}_{\mathbf{u}d_1} & \mathbf{K}_{\mathbf{u}d_2} & \mathbf{K}_{\mathbf{u}\beta} \\ \mathbf{K}_{\boldsymbol{\theta}\mathbf{u}} & \mathbf{K}_{\boldsymbol{\theta\theta}} & \mathbf{K}_{\boldsymbol{\theta}d_1} & \mathbf{K}_{\boldsymbol{\theta}d_2} & \mathbf{K}_{\boldsymbol{\theta}\beta} \\ \mathbf{K}_{d_1\mathbf{u}} & \mathbf{K}_{d_1\boldsymbol{\theta}} & \mathbf{K}_{d_1d_1} & \mathbf{K}_{d_1d_2} & \mathbf{K}_{d_1\beta} \\ \mathbf{K}_{d_2\mathbf{u}} & \mathbf{K}_{d_2\boldsymbol{\theta}} & \mathbf{K}_{d_2d_1} & \mathbf{K}_{d_2d_2} & \mathbf{K}_{d_2\beta} \\ \mathbf{K}_{\beta\mathbf{u}} & \mathbf{K}_{\beta\boldsymbol{\theta}} & \mathbf{K}_{\beta d_1} & \mathbf{K}_{\beta d_2} & \mathbf{K}_{\beta\beta} \end{bmatrix}, \qquad \mathbf{K}_{pq} = \mathbf{K}_{qp}^{\mathsf{T}} \qquad (30)$$

with $p, q$ ranging over the five field blocks, the symmetry following from Eq. (27) in every case. Two consequences for the block structure follow from Eq. (14). Differentiating $R_\beta$ with respect to the bulk fields gives $g_\beta{}'(\beta)\,g'(\varphi_f)\,g(\varphi_{if})\,\Psi_{int}^+$ and its counterpart in $\varphi_{if}$, so the bulk–interface blocks are nonzero wherever the interaction channel is fracture-active and the damage part of the tangent is block-dense rather than block-diagonal. Differentiating the interaction term twice with respect to the two bulk fields gives the cross-mechanism block $g'(\varphi_f)\,g'(\varphi_{if})\,g_\beta(\beta)\,\Psi_{int}^+$, which has no counterpart in a formulation whose channels are degraded independently. The Newton update is

$$\mathbf{K}(\mathbf{w}^{(i)})\,\Delta\mathbf{w}^{(i)} = -\mathbf{R}(\mathbf{w}^{(i)}), \qquad \mathbf{w}^{(i+1)} = \mathbf{w}^{(i)} + \Delta\mathbf{w}^{(i)} \qquad (31)$$

# 5. Finite element discretisation

## 5.1 Interpolation

Let $\mathcal{S}_0^h$ be a conforming partition of the shell midsurface into curved triangles $\Omega_e$. The kinematic fields are interpolated with quadratic Lagrange polynomials on the six-node triangle and the damage fields with linear Lagrange polynomials on its three vertices,

$$\mathbf{u}^h = \sum_{A=1}^{6} N_A^{(2)}\,\mathbf{u}_A, \qquad \boldsymbol{\theta}^h = \sum_{A=1}^{6} N_A^{(2)}\,\boldsymbol{\theta}_A, \qquad (d_1, d_2, \beta)^h = \sum_{a=1}^{3} N_a^{(1)}\,(d_1, d_2, \beta)_a \qquad (32)$$

giving thirty kinematic and nine damage unknowns, that is thirty-nine degrees of freedom per element. The unequal order is dictated by the constraints of Eq. (35), which are imposed at the nodes: a quadratic damage interpolation would admit nodal values satisfying the bounds while the interpolant violates them in the element interior.

The midsurface geometry carries the same quadratic basis, so that $b_{\alpha\beta}$, $H$ and $K$ are represented. A linear geometry basis annihilates the second derivatives of the position field and therefore returns $b_{\alpha\beta} = 0$ on every mesh, leaving the shifter of Eq. (2) inert; the quadratic basis recovers the curvature at first order, which is the expected rate for a second derivative of a quadratic map. For the benchmark geometries of Section 7.4 the chart is evaluated analytically, so that the closure residual of Eq. (9) is not contaminated by geometric interpolation error.

### 5.2 Quadrature

In-plane integration uses a degree-four Gauss–Jacobi rule with $n_\xi = 9$ points on the triangle. Through-thickness integration uses an $n_\zeta = 5$-station Gauss–Legendre rule on $[-h/2, h/2]$, the minimum that resolves the quadratic variation of the invariants induced by the shifter together with the interpolants of Eq. (12). The discrete mechanical energy is

$$\Pi^h_{mech} = \sum_e \sum_{q=1}^{n_\xi} \sum_{k=1}^{n_\zeta} w_q\ \omega_k\ \Psi_d\left(\mathbf{E}(\boldsymbol{\xi}_q, \zeta_k),\ d(\zeta_k)\right)\ J(\zeta_k)\ J_{\mathcal{S}}(\boldsymbol{\xi}_q) \qquad (33)$$

where $w_q$ and $\omega_k$ are the in-plane and through-thickness weights, $J_{\mathcal{S}} = \sqrt{\det\mathbf{G}}$ is the surface Jacobian and $J(\zeta_k)$ is the shifter of Eq. (2), which enters the discretisation only through the volume measure $\mathrm{d}V$ of Eq. (19). Because $\Pi^h$ is formed first and differentiated afterwards, $\mathbf{R}^h$ and $\mathbf{K}^h$ are the exact first and second variations of the same discrete functional.

The element is fully integrated: no reduced or selective quadrature, assumed- or enhanced-assumed-strain treatment, stabilisation term or shear correction factor is used. The kinematics of Eq. (3) carry no thickness-stretch degree of freedom, and plane stress is enforced constitutively through Eq. (16) rather than kinematically. The thin-limit behaviour of the resulting element is reported in Section 7.3.

### 5.3 Assembly and solution

Element and global operators follow by standard assembly,

$$\mathbf{R} = \sum_e \mathbf{A}_e^{\mathsf{T}}\ \mathbf{R}_e, \qquad \mathbf{K} = \sum_e \mathbf{A}_e^{\mathsf{T}}\ \mathbf{K}_e\ \mathbf{A}_e \qquad (34)$$

where $\mathbf{A}_e$ is the Boolean connectivity operator and $\mathbf{R}_e$, $\mathbf{K}_e$ integrate the residual and tangent densities of Section 4 over each element by the rules of Section 5.2.

The discrete system is solved by the Newton scheme of Eq. (31) subject to the pointwise irreversibility and box constraints

$$d_1^{n+1} \geq d_1^n, \qquad d_2^{n+1} \geq d_2^n, \qquad \beta^{n+1} \geq \beta^n, \qquad 0 \leq (d_1, d_2, \beta) \leq 1 \qquad (35)$$

relative to the last converged step $n$. The constraints are imposed by projection onto the feasible set rather than by penalisation, within an alternate-minimisation scheme whose damage block is solved by a projected Newton method. The linear systems are factorised with a sparse direct solver [56], whose pivot signs supply the shifted-inertia counts of Section 7.7.

## 6. Verification and certification

Three tiers organise what is checked before an admissible $\Psi_\alpha$, closed-form or learned, is used in a simulation. Each check is paired with a planted defect it must detect and with a correct state it must pass.

**Tier A — algebraic verification.** (V1) $\parallel \mathrm{FD}(\Pi) - \mathbf{R} \parallel < \tau_R$, evaluated block by block on each field rather than on the displacement block alone. (V2) $\parallel \mathrm{FD}(\mathbf{R}) - \mathbf{K} \parallel < \tau_K$ with $\mathbf{K} = \mathbf{K}_{mat} + \mathbf{K}_{geo}$ per Eq. (28), never $\mathbb{C}^{ps}$ alone. (V3) Finite-difference probes are flagged rather than scored as failures whenever the perturbation crosses one of the two switching surfaces $\{I_f = 0\}$ or $\{I_t = 0\}$ introduced by the Macaulay split of Eq. (11), as set out in Appendix A.7.

**Tier B — solver certification.** (S1) The accepted-equilibrium residual is measured on the complementarity object minimised by the active-set method rather than on plain stationarity, which a formulation carrying floor or feasibility terms could satisfy while remaining infeasible. (S2) The active set is unchanged after a classify–reset–solve cycle. (S3) Definiteness of the assembled tangent $\mathbf{K}$, not of $\mathbb{C}^{ps}$, is decided at the accepted state by the shifted-inertia count, the number of factorisation pivots below a stated floor. Because (S3) is a local, state-dependent signal that may legitimately fail near a limit state, the integer count is reported rather than the verdict alone.

**Tier C — learning certification.** A constitutive state enters a training set only if it passes both Tier A and Tier B at that state. This is a checkable admissibility criterion for data, orthogonal to whether the underlying law is analytic or learned.

Appendix A collects the definitions and proofs on which this section and Sections 3 and 4 rely.

## 7. Numerical results and discussion

The studies are ordered so that each relies only on what precedes it. Sections 7.1 and 7.2 fix the framework, the benchmarks and the admissibility of the deployed operator, the latter without a solver in the loop. Sections 7.3 and 7.4 verify the discretisation itself, in the thin regime and on curved geometry, before any quantity computed with it is interpreted. Sections 7.5–7.10 certify the discrete operators, the plane-stress condensation, the solver state, the crack functional, the mesh and the constitutive mapping. Sections 7.11–7.14 use the certified model to examine how fibre orientation, the interface field and curvature control fracture, and Section 7.15 summarises. The capacity of a learned representation seated in the same socket is measured separately in Appendix A.10. Table 1 lists the studies together with the control used in each case.

Table 1. Overview of the numerical studies. Each verification study is paired with a control that is capable of failing.

| **Fig.** | **Study** | **Quantity examined** | **Control** |
|---|---|---|---|
| 1 | Framework | Operator socket and admission gate | — |
| 2 | Benchmarks | Geometry, supports, material | — |
| 3 | Operator admissibility | Convexity, tangent eigenvalue, barriers | Certification threshold |
| 4 | Shell element, thin limit | $\Pi/t^3$ and its spread under refinement | Rotations in the per-cell basis |
| 5 | Curved-geometry chain | Flat limit and chart invariance | Spoofed geometry; planted metric |
| 6 | Residual and tangent | First and second variations of $\Pi$, all blocks | Tangent without $\mathbf{K}_{geo}$ |
| 7 | Plane-stress condensation | $S_{33}$ residual and $\partial S_{33}/\partial E_{33}$ | Condensation switched off |
| 8 | Tangent definiteness | Shifted-inertia count | Verdict floor on $\lVert \mathbf{K} \rVert$ |
| 9 | Toughness recovery | $G_{eff}/G_c$ against resolution ratio | Under-resolved mesh; $G_{IIc}$ |
| 10 | Mesh objectivity | Peak load and crack path | Coarsest rung |
| 11 | Constitutive correspondence | Mapping of a published energy | Deliberately wrong mapping |
| 12 | Fibre orientation | Damage envelope and driving forces | Matched peak-load state |
| 13 | Mechanism competition | Load capacity and driving shares | Seeded notch baseline |
| 14 | Interface field | Localisation and dissipation | Interface field frozen at zero |
| 15 | Curvature asymmetry | $\lVert d_u - d_l \rVert$ against $1/R$ | Flat panel at every step |
| A.1 | Distillation capacity | Learned operator error | Deployed network at zero |

### 7.1 Framework, benchmarks and material

Fig. 1 sets out the eight stages of the formulation and locates the constitutive operator among them. Stages 1–3 carry the shell kinematics of Eq. (3) to the Green–Lagrange strain of Eq. (5), to its Cholesky pull-back $\hat{\mathbf{E}}$ of Eq. (6), and to the invariant basis of Eq. (8), whose closure identity, Eq. (9), makes the mechanism decomposition exact. Stage 4 is the operator itself, drawn as a socket: the closed form is seated in it, the learned representation is admissible and certified but not deployed, and both pass the same Tier A/B/C admission gate. Stage 5 is the crack driving force of Eq. (17), placed between the operator and the degradation because the channels are signed. Stages 6–8 degrade the channels, assemble the regularised functional of Eq. (18) with the constraints of Eq. (35), and solve. Stages 5–8 are independent of the representation seated in stage 4; the two properties most at risk in an exchange of representation are proved in Appendices A.5 and A.6 and exercised in Appendix A.10.

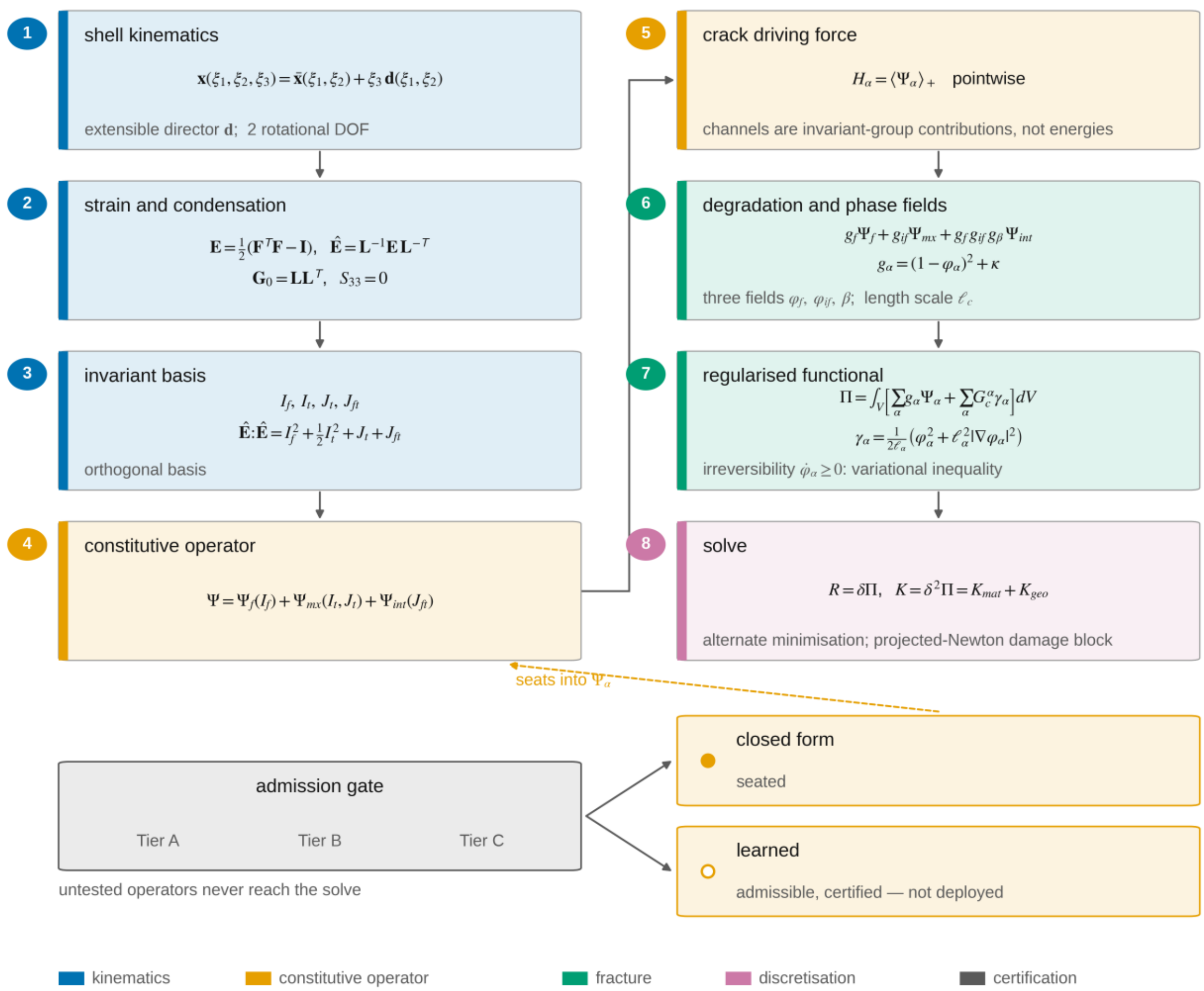

Fig. 1. The formulation in eight stages, read down the left column and then down the right. Stage 4 is the constitutive operator, into which the closed form is seated (filled marker) and the learned representation is admitted but not deployed (open marker). Both pass the same Tier A/B/C gate, and everything downstream of stage 4 is independent of the representation.

Fig. 2 defines the two benchmarks and the material. The single-edge-notched tension (SENT) specimen is 36 × 72 mm with a shell thickness of 1 mm and a notch $a_0 = 9$ mm at mid-height. The bottom row of nodes fixes all three displacement components and leaves the rotations free, and the top edge carries the prescribed displacement $\delta$. The fibre direction is swept over $\theta = 0$, 30, 60 and 90°, measured from the horizontal, so that $\theta = 90°$ places the fibres parallel to the loading direction and $\theta = 0°$ perpendicular to it. The clamped bending specimen shares the plate dimensions, carries its notch 6 mm from the clamped edge, fixes displacement and rotation at the wall, and is loaded out of plane at the tip.

The property table of Fig. 2(c) separates the measured ply constants from the invariant coefficients derived from them, since the latter are what the operator consumes. The relations that produce them, $\mu_{LT} = 2(G_{LT} - G_{TT}) + 2\mu_{mII}$ with $G_{TT} = E_T/[2(1 + \nu_{TT})] = 3.931$ GPa, are printed with the table so that the mapping from a datasheet to the invariant basis can be reproduced. The shell thickness $h$ and the element size $h_e$ are listed separately although both equal 1 mm on the SENT mesh, because they enter different expressions: the thickness converts the through-thickness integral of Eq. (33) from an areal density in J/m² to a volumetric density in J/m³, whereas the element size enters only the resolution ratio $\ell_c/h_e$. The fibre toughness exceeds the inter-fibre toughness by a factor of 1630, and this ratio governs the orientation dependence reported in Section 7.12.

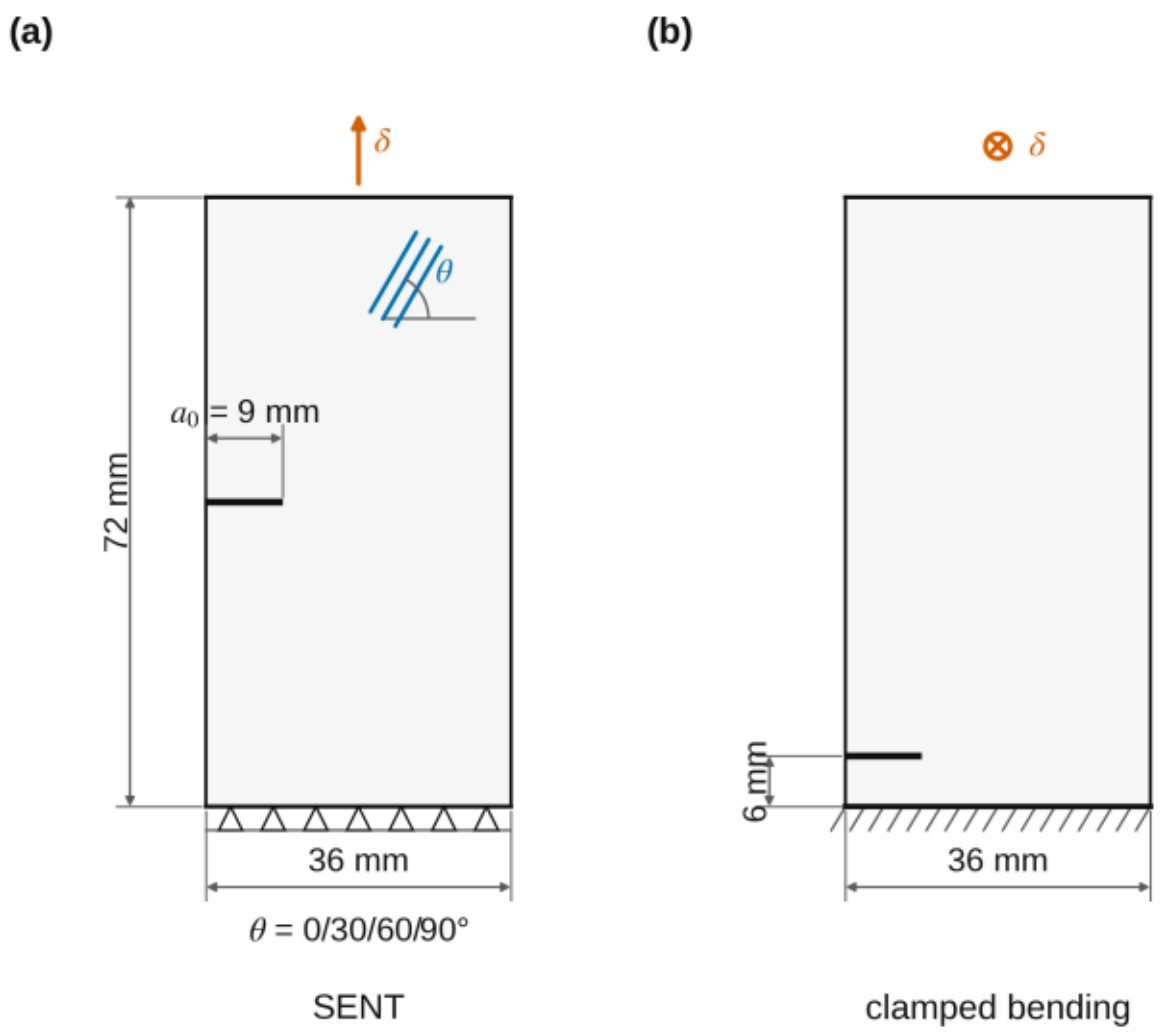

(c)

| quantity | symbol | value | unit |
|---|---|---|---|
| *ply constants (measured)* | | | |
| longitudinal modulus | $E_L$ | 161 | GPa |
| transverse modulus | $E_T$ | 11.4 | GPa |
| major Poisson ratio | $\nu_{LT}$ | 0.32 | – |
| transverse Poisson ratio | $\nu_{TT}$ | 0.45 | – |
| in-plane shear modulus | $G_{LT}$ | 5.17 | GPa |
| *invariant coefficients (derived from the ply constants)* | | | |
| transverse-opening coefficient | $\lambda_{mI}$ | 3.635 | GPa |
| transverse-shear coefficient | $\mu_{mII}$ | 3.735 | GPa |
| LT-shear coefficient | $\mu_{LT}$ | 9.947 | GPa |
| star-convex exponent, TI | $\gamma_T$ | 6.74 | – |
| star-convex exponent, isotropic | $\gamma^*$ | 0 | – |
| *fracture* | | | |
| matrix toughness | $G_c$ | 50 | J/m² |
| fibre toughness | $G_c^f$ | 81.5 | kJ/m² |
| interface toughness, swept | $G_c^{int}$ | 10–1000 | J/m² |
| *length scales* | | | |
| regularisation length | $\ell_c$ | 5 | mm |
| shell thickness | $h$ | 1 | mm |
| element size | $h_e$ | 1 | mm |

$\mu_{LT} = 2(G_{LT} - G_{TT}) + 2\mu_{mII}$, $G_{TT} = E_T/2(1+\nu_{TT}) = 3.931$ GPa

Fig. 2. Benchmark definition. (a) Single-edge-notched tension specimen, $36 \times 72$ mm, thickness 1 mm, notch $a_0 = 9$ mm; the pin row fixes displacement and leaves rotation free, and the rosette gives the fibre direction. (b) Clamped bending specimen at the same scale, with the tip displacement out of plane. (c) IM7/8552 properties used by every solve, with the invariant coefficients derived from the measured ply constants.

## 7.2 Admissibility of the deployed operator

Before the operator enters a boundary-value problem it is evaluated pointwise, with no solver in the loop, against the conditions of Definition A.1. Fig. 3(a) grows each channel along a straight ray in the invariant that excites it, namely $I_f$ for the fibre channel, $(I_t, J_t)$ for the inter-fibre channel and $J_{ft}$ for the interaction channel, so that each ray probes one channel while the others remain at the origin. The response is quadratic on all three rays, the smallest second difference along any ray is positive, and the two branches of each ray agree to $3.2 \times 10^{-6}$, so the Macaulay split of Eq. (11) introduces no asymmetry where none is intended. Fitting the curvatures returns the operator coefficients $\lambda_f$, $2\lambda_{mI}$ and $\mu_{LT}$ to $4.0 \times 10^{-4}$ relative. This checks the map from invariants to coefficients rather than the plausibility of the output, since a transposition or a missing factor of two in that map would leave the curves quadratic and shift the fitted curvature beyond the stated tolerance.

Fig. 3(b) reports the magnitude of the smallest eigenvalue of the $6 \times 6$ material tangent, normalised by the tangent norm, over 203 sampled states. Every point satisfies $|\lambda_{\min}(\mathbb{C})|/\| \mathbb{C} \| < 4 \times 10^{-16}$, eight decades under the certification threshold of $10^{-8}$. The normalisation is required rather than cosmetic: each channel depends only on its own invariants, so its Hessian is rank-deficient by construction and the exact zero eigenvalue appears numerically as a small negative number whose absolute magnitude scales with the tangent norm. For the fibre and inter-fibre channels, convexity in $\hat{\mathbf{E}}$ is proved in Appendix A.2 and, for the two linear invariants, strengthened to convexity in $\mathbf{F}$ directly in Appendix A.3. For the interaction channel the corresponding statement in $\mathbf{F}$ is left open in Appendix A.4, and the pointwise certificate of panel (b) stands in its place.

**Remark 4.** A pointwise certificate is not a proof: it reports only on the states sampled. Fig. 3(b) therefore marks the fibre and inter-fibre channels as proved and the interaction channel as certified.

Fig. 3(c) shows the two coercivity barriers as their arguments approach the boundary of admissibility. They supply condition (A5), which convexity does not imply [55], and they are reported as themselves: they are never folded into a mechanism contribution and never degraded, since degrading a barrier by $g(d)$ would leave it finite at full damage and the constraint it enforces would disappear where the deformation is least controlled. At the reference state the operator is exactly stress-free, with $\Psi_\alpha = 0$ and $\max \| \mathbf{S} \| = 0$ for all three channels, which is condition (A3) and is what makes the local coercivity estimate of Eq. (A.7) available.

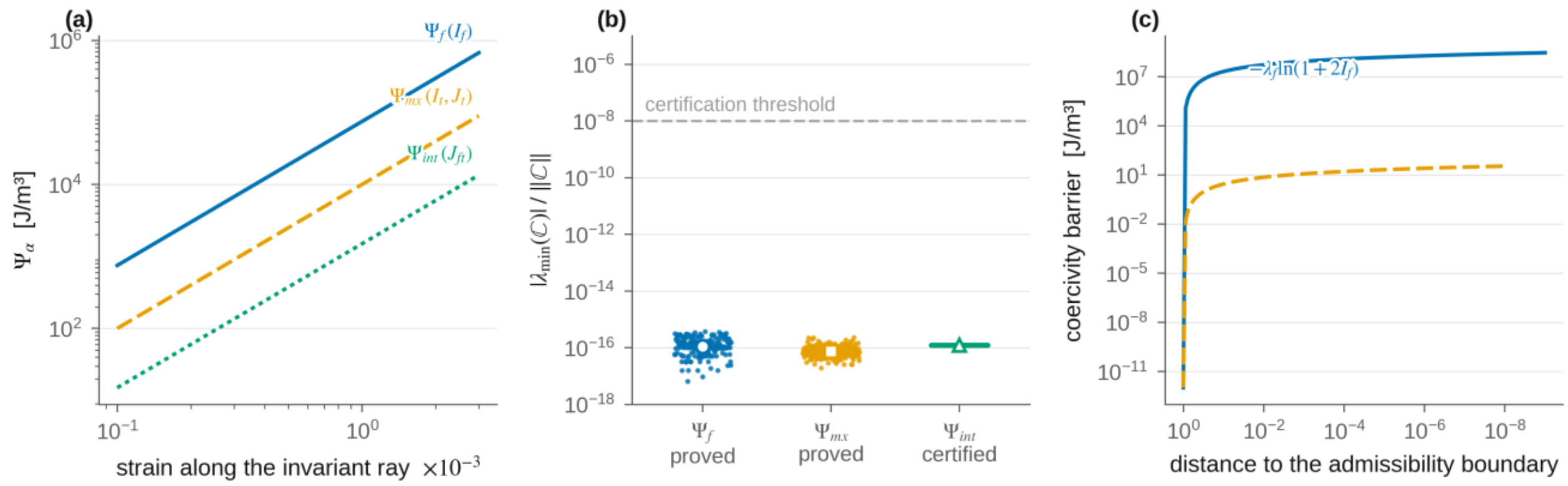

Fig. 3. Pointwise evaluation of the deployed operator, without a solver. (a) Growth along one straight ray per mechanism, un-normalised. (b) Magnitude of the smallest eigenvalue of the $6 \times 6$ material tangent relative to the tangent norm over 203 sampled states, against the certification threshold of $10^{-8}$; channels are labelled as proved or certified. (c) The two coercivity barriers as their arguments approach the admissibility boundary.

### 7.3 The shell element in the thin limit

Section 5 states the element without any assumed-strain or reduced-integration device, so its behaviour as the shell thins must be established before any fracture result computed on it is interpreted. Fig. 4 reports that measurement on an exact inextensional deformation of the skew cylinder, driven on the deployed element. The test is chosen because an inextensional field should cost bending energy alone, so that a parasitic membrane or transverse-shear contribution appears as an excess that grows as the thickness falls, which is the locking signature.

Panel (a) reports the strain energy scaled by $t^3$ over a sixteen-fold range of thickness $t$, so that the thin-plate bending law is a horizontal line. The deployed element follows $\Pi \sim t^{2.937}$, within 2.1% of the exact cubic law across the whole ladder. The control is the same element with the director rotations taken in the per-cell basis rather than a common frame, retained as a fail-capable reference; it follows $\Pi \sim t^{1.037}$, four orders above the deployed curve and steepening as the shell thins.

Panel (b) is the discriminator between locking and discretisation error. Locking is a property of the interpolation space and therefore survives mesh refinement, whereas discretisation error does not. The spread of $\Pi / t^3$ across the thickness ladder falls from 1.61 to 1.20 for the deployed element as the mesh is refined from $8 \times 8$ to $16 \times 16$, moving towards the exact law, while for the planted defect it rises from 217.2 to 231.8, moving away from it. The two legs move in opposite directions under the same refinement, which is a difference in kind rather than in degree. Panel (c) repeats the measurement against radius at fixed $t = 0.50$ mm and recovers $\Pi \sim R^{-2.003}$ against the exact $R^{-2}$ law, with $\Pi R^2$ constant to 1.0084 over a sixteen-fold range of radius.

The quadratic kinematic interpolation of Eq. (32) therefore does not lock over the slenderness range examined, and no assumed-strain treatment is required for the results reported here. Both parasitic laws fit the deployed data with negative coefficients, so no parasitic contribution can be attributed to membrane or to transverse shear separately, and none is claimed.

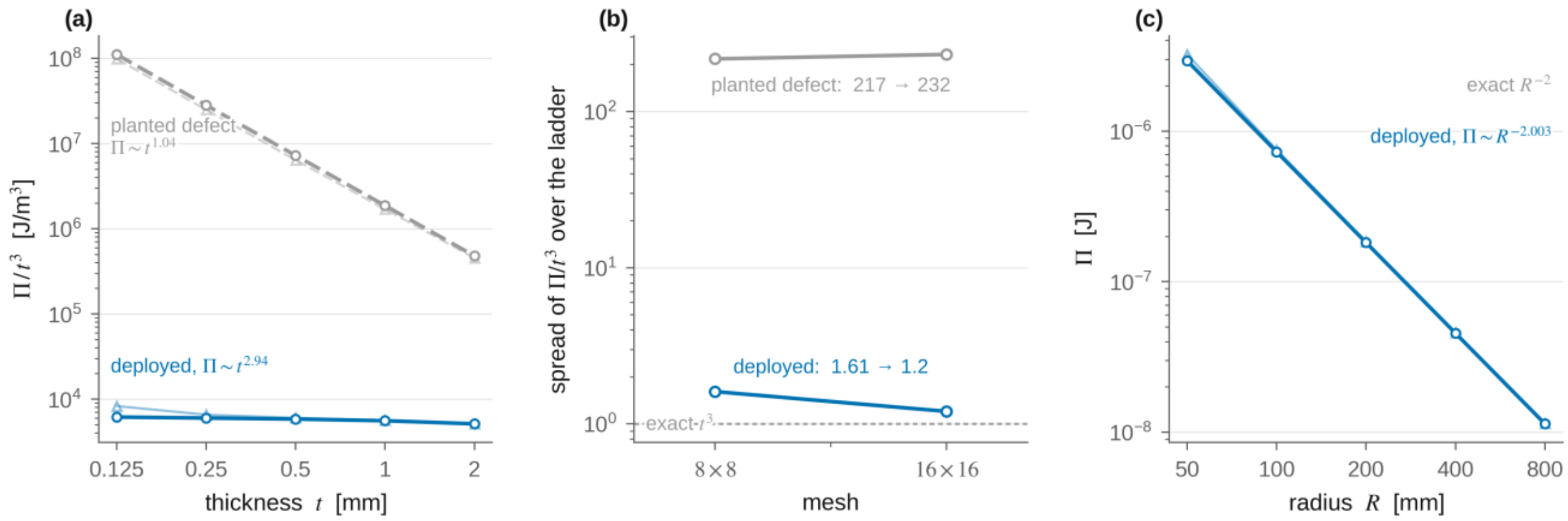

Fig. 4. Locking test on an exact inextensional deformation of the skew cylinder, driven on the deployed element. Faint triangles are the $8 \times 8$ mesh and solid circles the $16 \times 16$ mesh. (a) Strain energy scaled by $t^3$ over a sixteen-fold thickness range, so that the bending law is a horizontal line. (b) Spread of $\Pi/t^3$ across the ladder against mesh, the discriminator between locking and discretisation error. (c) The same energy against radius at $t = 0.50$ mm, against the exact $R^{-2}$ law. The planted defect is the same element with the rotations taken in the per-cell basis and is a fail-capable reference.

### 7.4 Certification of the curved-geometry chain

Every study reported to this point is carried out on a flat midsurface, where $b_{\alpha\beta} = 0$, the shifter of Eq. (2) is the identity and the metric machinery of Eqs. (6)–(9) is exercised only in its non-orthogonal aspect. Fig. 5 certifies the same chain on curved geometry, where the shifter, the thickness dependence of $\mathbf{L}(\zeta)$ and the through-thickness variation of the invariants are all active. Each panel is paired with a control constructed to fail.

Panel (a) reports the relative departure of the curved solution from the flat plate against radius. The deployed chain converges monotonically at order 1.72 in $1/R$, from 0.926 at $R = 0.1$ m to $2.15 \times 10^{-7}$ at $R = 1000$ m, so the flat solution is recovered as the curvature is removed. The control is a spoofed geometry that carries the curved label without its metric; it returns the flat answer at every radius, which is what curvature failing to enter the kinematics looks like.

Panel (b) tests chart invariance, which the flat studies cannot pose. The four invariants of Eq. (8) are evaluated for one physical state described in three charts of the same surface, one natural and two skewed, and agree to at most $2.5 \times 10^{-15}$. The upper references are quantities of the same states that are not chart-invariant: the metric and curvature components $G_{12}$ and $b_{11}$, which differ by order unity between the charts, and a planted metric substituted into the pull-back, which deviates by 0.743. Fifteen orders of magnitude separate what is invariant from what is not. The distinction matters beyond verification, because it is what licenses reporting curvature through $H$ and $K$ rather than through components of $b_{\alpha\beta}$, which take different values on different charts of the same surface.

Panel (c) shows the through-thickness profile of the four invariants at $R = 100$ mm as a percentage of their midsurface value. The invariants vary across the thickness by between two and three per cent, which is curvature entering the kinematics rather than being carried by the midsurface alone, and is the mechanism behind the asymmetry reported in Section 7.14. The variation tracks the shifter: it is largest for the invariants that sample the direction of nonzero curvature and is absent along the generator, where $\mathbf{b}$ vanishes.

One limitation of panel (b) is stated rather than left implicit. The closure identity of Eq. (9) tests the normalisation of the pull-back and nothing else, so it is blind to a consistent substitution of the metric; the invariance demonstrated here is invariance under change of chart, not a proof that the pull-back is the intended one. The closure residual is reported in Table 1 and is not evidence on its own.

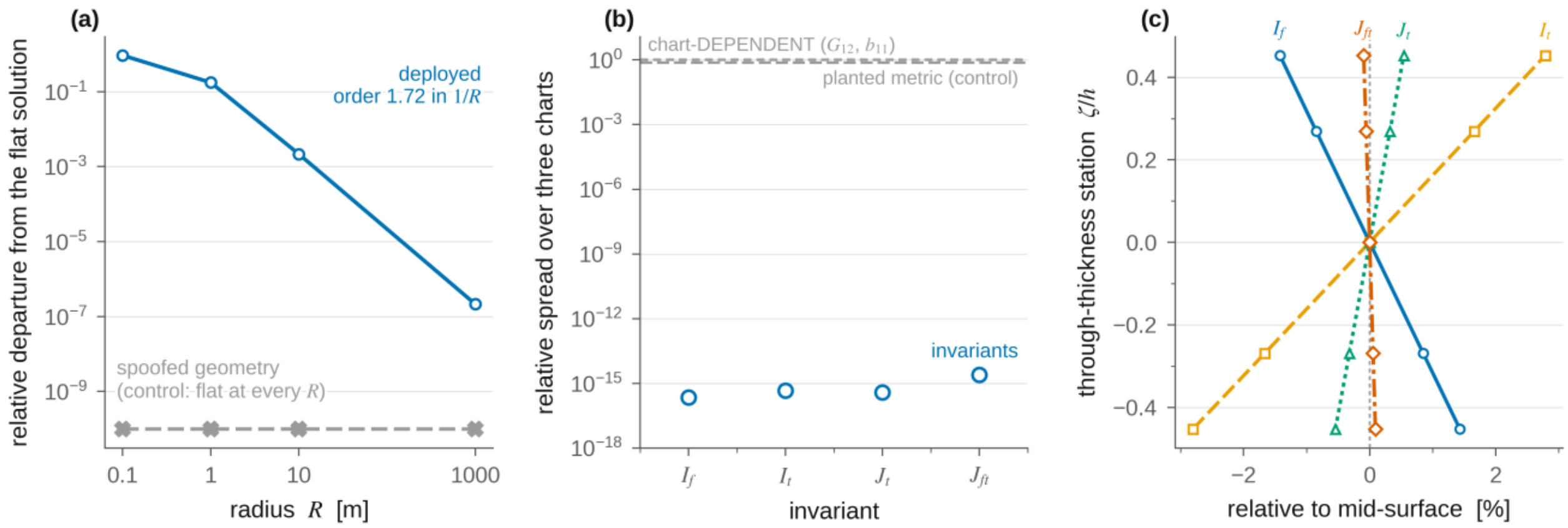

Fig. 5. Certification of the curved-geometry chain, driven on the deployed element. (a) Relative departure of the curved solution from the flat plate against radius, with a spoofed geometry that carries the curved label without its metric as control. (b) Relative spread of each invariant across three charts of the same surface, against chart-dependent quantities and a planted metric. (c) Through-thickness profile of the four invariants at $R = 100$ mm as a percentage of their midsurface value.

### 7.5 Residual and tangent as variations of a single potential

Tier A asks whether **R** and **K** are the first and second variations of the same $\Pi$, which is the only guarantee that Newton's method solves the problem that was posed. Fig. 6 answers this by central differences at a loaded, converged state, driven on the element kernel over 24 displacement probes. In panel (a) the difference between the numerical first variation of $\Pi$ and the assembled residual falls with slope $h^2$ to $2.7 \times 10^{-10}$ and then rises again. Both branches are informative: the descending branch is the truncation error of a central difference, whose slope identifies the derivative as exact, while the ascending branch is cancellation, in which the difference of two nearly equal energies loses significant digits at a rate $h^{-1}$. A derivative that were only close to correct would produce a floor at the level of the discrepancy rather than a minimum.

Panel (b) carries the control. The full tangent $\mathbf{K} = \mathbf{K}_{mat} + \mathbf{K}_{geo}$ converges to $1.4 \times 10^{-11}$, again with slope $h^2$, whereas the same probes at the same state with the geometric term omitted stall at $1.4 \times 10^{-5}$ and do not improve as the perturbation is reduced. This is the numerical statement of Appendix A.6. The director of Eq. (3) is an extensible vectorial update, so **F** is affine in the unknowns, $D^2\mathbf{F} = \mathbf{0}$ exactly, and the second variation of **E** reduces to the Gram term of Eq. (A.5). That term is generically nonzero and depends on **S** and the kinematic operator alone, never on $\mathbb{C}$. Exchanging the constitutive representation therefore rescales $\mathbf{K}_{mat}$ and leaves $\mathbf{K}_{geo}$ unchanged, so a tangent assembled from $\mathbb{C}^{ps}$ alone is wrong for every admissible operator by the same six decades.

**Remark 5.** At the load levels considered here the geometric term is small enough that the Newton iteration still converges quadratically without it, so convergence of the iteration is not by itself evidence of a correct tangent.

The shaded band marks perturbations in which every probe steps across a constitutive kink. The Macaulay split of Eq. (11) composes each channel with a ramp applied to the linear switching functionals $I_f$ and $I_t$, so $\Psi_d$ is only piecewise smooth across $\{I_f = 0\} \cup \{I_t = 0\}$, as set out in Appendix A.7. A difference quotient taken across such a surface compares two different tangents and its value is set by the size of the jump; the kernel detects the crossing through its own witness field, and these probes are excluded rather than scored as failures.

Panel (c) checks the regularisation. The fibre phase field measured normal to the crack is compared with the analytic AT2 profile $\exp(-|s|/\ell_c)$ at three load levels on the finest mesh, $\ell_c/h_e = 8.3$. The maximum deviation grows from 0.033 to 0.085 as the wake lengthens, which is the expected direction, since the analytic profile is the one-dimensional optimal solution for a straight crack in an infinite body. No parameter is fitted to the reference.

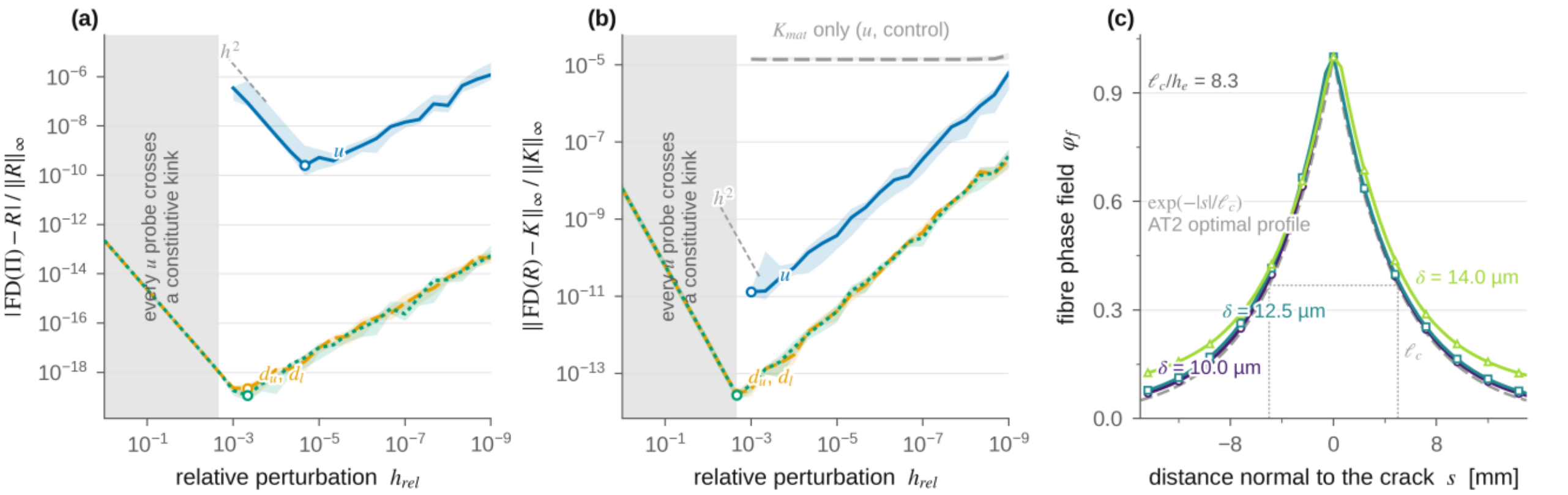

Fig. 6. Central-difference verification at a loaded, converged state (step 27, $\delta = 18.90$ µm, $30 \times 60$ mesh), driven on the element kernel. Each field block is perturbed in its own degrees of freedom and compared against its own residual entry and its own diagonal tangent sub-block on one relative ladder $h_j = h_{rel}\max(|x_j|, x_{scale})$; $\mathbf{u}$, $d_u$ and $d_l$ are active at this state and are drawn, and the remaining blocks are reported in the accompanying table rather than on an axis they do not share. Solid lines are medians, bands the interquartile range, and dashed guides have slope $h^2$. (a) First variation: every block falls to $2.7 \times 10^{-10}$ or below. (b) Second variation: the full tangent reaches $1.4 \times 10^{-11}$ while the material-only control, at the identical state with the geometric term omitted, stalls at $1.4 \times 10^{-5}$ and never converges. The shaded band marks the perturbation range in which every displacement probe crosses a constitutive kink, detected from the kernel's own witness and excluded rather than plotted; the damage blocks cross no kink at any rung, which is why their curves continue through it. (c) Fibre phase field normal to the crack against the analytic AT2 profile.

### 7.6 Plane-stress condensation

The shell response is plane-stress, enforced by solving $S_{33}(E_{33}) = 0$ locally and condensing on the 33-component as in Eq. (16), so the accuracy of that local solve propagates into every stress and every tangent. Fig. 7 measures it along the bending load path at 233,280 quadrature points per state. In panel (a) the condensed kernel holds $|S_{33}|$ at about $4 \times 10^{-24}$ of the in-plane stress scale, which is the arithmetic floor for the stress magnitudes involved rather than a prescribed tolerance, while the same states with the condensation switched off sit at about $10^{-7}$. The seventeen-decade separation is what makes the uncondensed curve a usable control.

Panel (b) compares the thickness strain recovered by an independent Newton solve on $S_{33} = 0$ with the value supplied to the kernel. The recovered strain grows with load and remains bounded, Newton converged at every sampled point, and the Schur condensation was never skipped at any load, including the states beyond compression-surface cracking marked by the dashed line. Because the check uses a different solver on the same equation, agreement is evidence about the kernel rather than about the iteration.

Panel (c) certifies the hypothesis of Appendix A.5 rather than its conclusion. The implicit function theorem preserves the regularity of $E_{33}^*$ only where $\mathbb{C}_{3333} = \partial S_{33}/\partial E_{33} \neq 0$, and the appendix shows that the residual stiffness $k$ of Eq. (13) keeps that derivative bounded away from zero even at full damage, since $g(d) \geq k > 0$ throughout. The measurement returns a minimum of 3.244 GPa over 400 quadrature points across five load steps taken to $d = 0.77$, with no non-positive value. The residual stiffness introduced for numerical safety therefore performs a second, structural function. A consequence for practice is that $k$ cannot be reduced without limit in order to sharpen the softening response, and panel (c) is the diagnostic that would register the loss first.

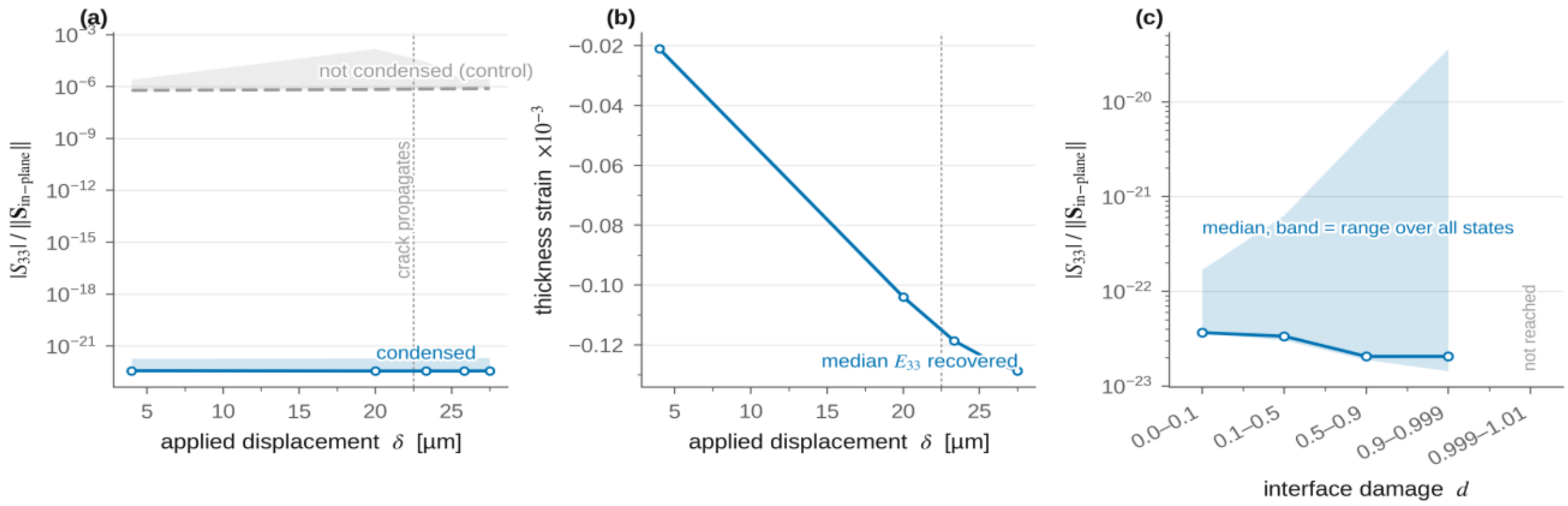

Fig. 7. Condensation diagnostics along the bending load path, sampled at 233,280 quadrature points per state. (a) Through-thickness stress relative to the in-plane stress scale, for the condensed kernel (median and 99th-percentile band) and for the same states with the condensation switched off. (b) Thickness strain recovered by an independent Newton solve on $S_{33} = 0$ against the value supplied to the kernel; the dashed line marks compression-surface cracking. (c) $\partial S_{33} / \partial E_{33}$ along the same path.

## 7.7 Definiteness of the assembled tangent

Tier B asks a different question from Tier A, namely whether the state at which the tangent is evaluated is a minimiser. Fig. 8 answers this for the $36 \times 72$, $\theta = 60°$ run at 28 states along the equilibrium path, on the free-degree-of-freedom tangent with 60,590 unknowns. The smallest eigenvalue is obtained by bisection on the shifted-inertia count: the matrix $\mathbf{K} - \sigma\mathbf{I}$ is factorised for a trial shift $\sigma$, and the number of negative pivots returned by the $\mathbf{LDL}^{\mathrm{T}}$ factorisation [56] gives, by Sylvester's law of inertia, the exact number of eigenvalues below $\sigma$. About 26 factorisations per state bracket the smallest eigenvalue to the width plotted.

No eigensolver is used, and the reason is structural. A softening tangent has no spectral gap near zero, since the damage block contributes a dense cluster of small eigenvalues as $g(d)$ approaches $k$, so a Krylov or LOBPCG estimate of $\lambda_{\min}$ depends on the stopping criterion and may be reported as negative when it is not, or as positive when it is not. A pivot-sign count has no such failure mode, because it counts exact integers and answers a well-posed question, namely how many eigenvalues lie below a stated shift.

With the verdict floor at $\sigma = -10^{-10} \parallel \mathbf{K} \parallel$, the tangent is positive semi-definite through the peak and along most of the softening branch, and loses definiteness only at the last two states, where the counts are exactly 3 and then 4 negative modes. Definiteness therefore survives to 14% of the peak load on the descending branch. This is the expected behaviour of a fully softened tangent rather than a solver failure, and it is anticipated by criterion (S3) of Section 6. Reporting the count rather than only the verdict makes the distinction usable, since three negative modes at the deep end of softening is a different statement from an indefinite tangent at the peak, and the count also bounds the dimension of the subspace that a continuation scheme would have to control.

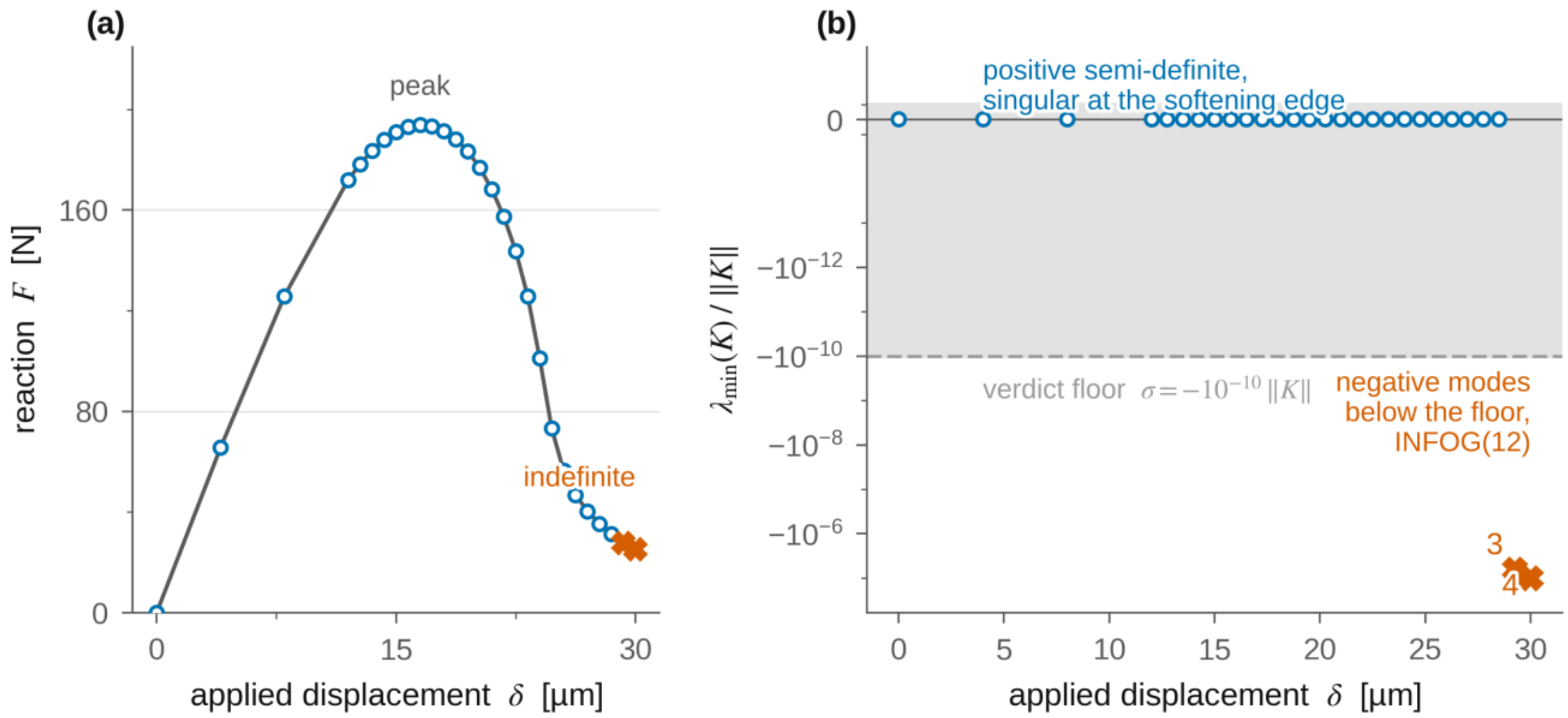

Fig. 8. Definiteness of the assembled tangent along the equilibrium path of the $36 \times 72$, $\theta = 60°$ run. (a) Load–displacement response with the 28 states at which the tangent was reassembled and factorised; circles denote positive semi-definite states and crosses indefinite ones. (b) Smallest eigenvalue of the free-degree-of-freedom tangent (60,590 unknowns), drawn as the interval returned by bisection on the shifted-inertia count, with the verdict floor $\sigma = -10^{-10} \parallel \mathbf{K} \parallel$ and the exact number of eigenvalues below it.

### 7.8 Recovery of the prescribed toughness

The regularised functional of Eq. (20) is useful only if the energy it dissipates equals the toughness it was given. Fig. 9 reports the ratio $G_{eff}/G_c$ for a prescribed value of 969 J/m², taken from the interlaminar data of Camanho et al. [2], as its deviation from unity against the resolution ratio $\ell_c/h$. The ladder is arranged so that two entries share each ratio while their length scale and mesh size differ by a factor of two. Those pairs coincide to six figures, giving 1.0066 at $\ell_c/h = 2.5$ and 1.0017 at $\ell_c/h = 5$, which localises the residual error in the resolution ratio alone rather than in either quantity separately. A ladder in which only the mesh were refined would show convergence but could not separate a discretisation error from a length-scale effect.

Two controls are drawn as crosses. An under-resolved mesh at $\ell_c/h = 0.42$ misses by 21.5%, which confirms that the diagnostic responds to under-resolution rather than passing by construction, and prescribing the mode-II toughness in place of the mode-I value is rejected at 43.3%, which shows that the recovered quantity tracks the toughness actually supplied. The dissipated energy scales exactly with the prescribed toughness, returning a ratio of 2 for a factor-of-two change, as the linear dependence of Eq. (20) on $G_c$ requires.

**Remark 6.** Under the surface-resolved parameterisation of Section 3.3, a crack running through the full thickness activates both surface fields and is charged 2.013 $G_c$. Any identification of this $G_c$ with a measured toughness must carry that factor. What is verified here is the crack functional as implemented; delamination is not validated.

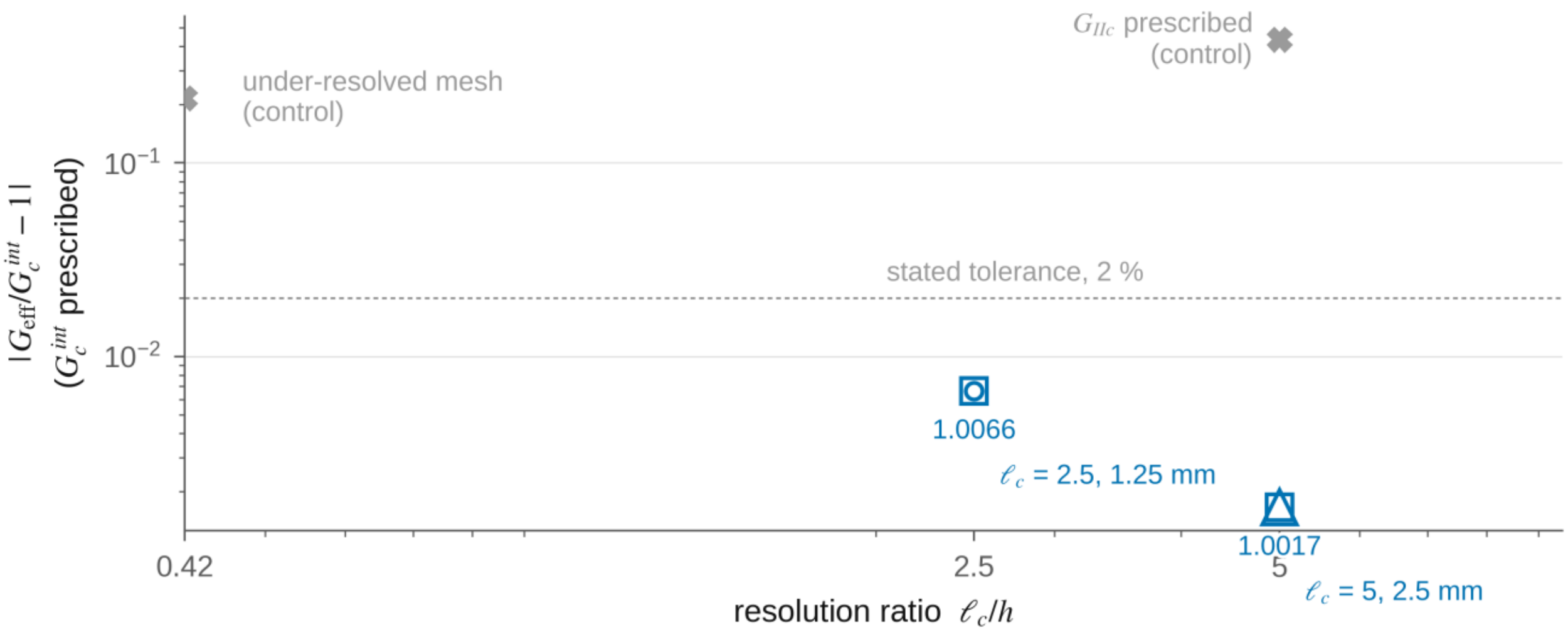

Fig. 9. Recovery of the prescribed toughness by the implemented AT2 surface energy at $G_c = 969$ J/m² [2], shown as the deviation of $G_{eff}/G_c$ from unity against the resolution ratio $\ell_c/h$. Two ladder entries share each ratio although their $\ell_c$ and mesh size differ by a factor of two. The crosses are controls: an under-resolved mesh, and the mode-II toughness prescribed in place of the mode-I value.

## 7.9 Mesh objectivity

Fig. 10 refines the SENT mesh over four rungs, from $30 \times 60$ to $60 \times 120$ elements, giving element sizes of 1.20, 1.00, 0.75 and 0.60 mm and resolution ratios $\ell_c/h_e$ from 4.2 to 8.3. The load–displacement curves in panel (a) are indistinguishable up to the peak and separate only in the steepest part of the descending branch, where the response is controlled by how well the localised band is resolved and where a difference of one element across the band changes the dissipated energy per increment. The peak reaction falls monotonically from 363.5 N to 356.1 N and changes by 0.20% between the two finest rungs, so the sequence has reached a plateau. Monotone approach from above is expected, since a coarse mesh over-resolves the stiffness of the band.

Panel (c) tracks the crack-tip abscissa on the notch line and panels (d)–(g) show the damage field at the last converged step of each rung on one shared scale. The band narrows as the element size falls, which is the expected behaviour of a fixed $\ell_c$ resolved by more elements, while the path itself does not move. This distinction is what mesh objectivity means for a regularised model: the length scale is a material parameter and the mesh must resolve it, so convergence means a path and a peak that stop changing under refinement rather than a band whose width stops changing. A band width that followed the element size would indicate the opposite, namely that the regularisation had been lost and the localisation had become mesh-dependent.

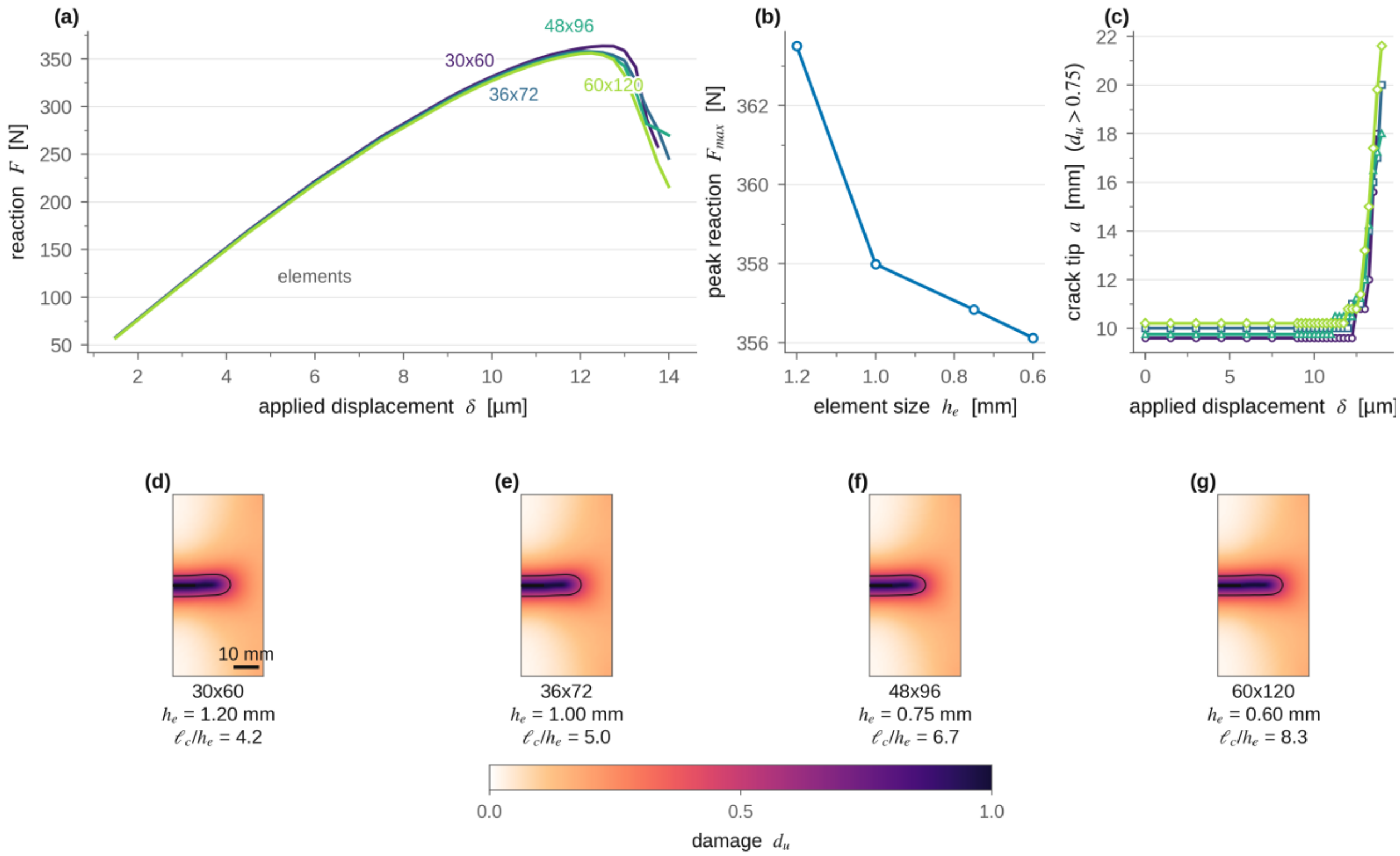

Fig. 10. Mesh ladder on the SENT specimen, four rungs from 30 × 60 to 60 × 120 elements ($h_e = 1.20$ to 0.60 mm). (a) Reaction against applied displacement. (b) Peak reaction against element size. (c) Crack-tip abscissa on the notch line, $d_u > 0.75$. (d)–(g) Damage field at the last converged step of each rung on one shared scale. Verification material; discretisation study only.

## 7.10 Correspondence with a published anisotropic model

The invariant basis of Eq. (8) is a change of variables, so an established anisotropic energy written on a different basis must map into it exactly. Fig. 11 performs this test against Eq. (129) of Teichtmeister et al. [26], using their geometry, boundary conditions and material, with the fibre at 45°. Panel (d) reports the worst relative residual between the two energies over the sampled state space as $3.4 \times 10^{-16}$, in plane strain and in full three dimensions, against a deliberately wrong mapping that misses by $5.2 \times 10^{-3}$. The thirteen-decade gap makes the agreement a statement about the algebra rather than about tolerance. Evaluating in three dimensions as well as in plane strain matters here, because a mapping error confined to the out-of-plane components would survive a plane-strain check unnoticed.

The crack paths in panels (a) and (b) run straight ahead of the notch rather than along the fibre. In the reference, path anisotropy enters the crack surface density through a term $\alpha_s(\nabla d \cdot \mathbf{m})^2$, which the present formulation does not carry, so the model sits at $\alpha_s \equiv 0$ by construction. The energy anisotropy alone is weak for this material, $\chi/\mu = 3.3\%$, which is consistent with the straight paths observed and indicates that the crack turning reported in the reference is carried by the surface-density term rather than by the stored energy. Panel (c) reports the length-scale dependence instead: halving $\ell_c$ raises the peak by a factor of 1.319, against the $\sqrt{2} = 1.414$ expected of an AT2 model in which the nucleation stress scales as $\ell_c^{-1/2}$. The shortfall is the finite-size correction expected when the process zone is not small compared with the ligament, and it confirms that the peak is $\ell_c$-controlled rather than strength-controlled.

**Remark 7.** No crack-path comparison is attempted. The reference publishes trajectories only at $\alpha_s \in \{-0.5, 1, 5, 50\}$, so no published path exists at the single parameter value that this formulation can reach. Their regularisation length of 0.075 mm would in addition require about 143,000 cells, so the peak load is not presented as a quantitative match either.

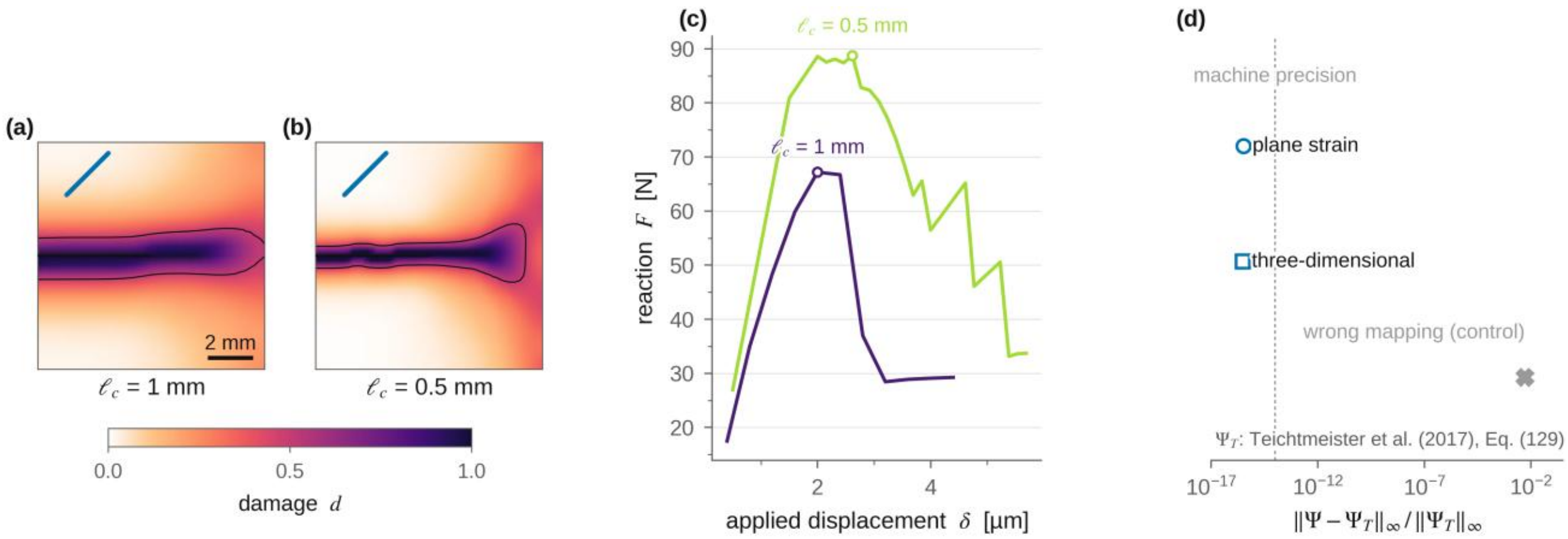

Fig. 11. Constitutive correspondence with Teichtmeister et al. [26], using their geometry, boundary conditions and material, with the fibre at 45°. (a, b) Damage field at two regularisation lengths. (c) Load–displacement for both lengths. (d) Their Eq. (129) mapped into the present invariant basis, against a wrong-mapping control.

## 7.11 Fibre orientation and the damage envelope

With the formulation certified, the fibre angle is swept. Fig. 12 compares the four solves at the matched state $\delta/\delta_{peak} = 1.00$, which is the only ratio all four reach, since the $\theta = 90°$ solve is arrested at its own limit point. Comparison at a matched fraction of the peak rather than at matched displacement is necessary because the peak displacements differ by a factor of seven, from 16.5 µm at 60° to 115.7 µm at 90°; at matched displacement the four fields would be at four different points of their load histories.

Panels (a)–(d) show the inter-fibre damage $\varphi_{if}$. The separated material, drawn white over the diffuse field, stays at the notch at every angle, and panel (e) shows why the distinction matters: the area above $\varphi_{if} = 0.5$ varies by a factor of 7.0 over 0–90°, while the area above $\varphi_{if} = 0.9$ remains between roughly 9 and 19 mm². Orientation therefore controls the envelope of diffuse damage and the load at which it forms rather than the direction the crack takes. Where enough material has grown beyond the seed for a principal axis to be meaningful, taken here as at least 100 nodes, that axis follows the fibres: 59.8° against a 60° fibre direction from 663 grown nodes, and 90.3° against 90° from 625 nodes. At 0° and 30° only 36 and 86 nodes grow, which is too few to define an orientation, and none is claimed. The threshold is stated because a principal axis computed from a nearly circular set is dominated by the seed geometry.

Panels (f)–(i) show the inter-fibre crack driving force $H_{if}$ of Eq. (17) on a shared logarithmic scale, with the $\varphi_{if} = 0.5$ contour drawn over it so that damage and its driver are read together. At 0, 30 and 60° the driving force concentrates at the notch tip and decays into the body, which is the distribution expected of a notch-tip field in a material whose weak direction is loaded. At 90° the pattern inverts and a large part of the specimen sits at the floor of the scale, which is the shielding of Section 3.2 made visible. With the fibres parallel to the load, the axial strain is carried almost entirely by the fibres and the transverse strain is contractive over most of the specimen, so $I_t < 0$ engages the star-convex term, the inter-fibre contribution is driven negative at 45.5% of the quadrature points, and the positive part in Eq. (17) vanishes there. The shielded region is not a region of low stress; it is a region in which the transverse mechanism has been made deliberately insensitive to a compressive transverse state.

Panel (j) collects the specimen-integrated driving forces at the matched state. All three are nonnegative at every angle, which is why the axis can be logarithmic and the series continuous; the same plot for the signed channel contributions could not be drawn at 90°. The fibre driving force rises by roughly six decades from 0 to 90° while the inter-fibre and interaction forces vary within about one decade, which is the quantitative form of the mechanism transfer examined in Section 7.12. The $\theta = 90°$ values are lower bounds, drawn with arrows, because that solve is arrested at a limit point rather than carried through softening.

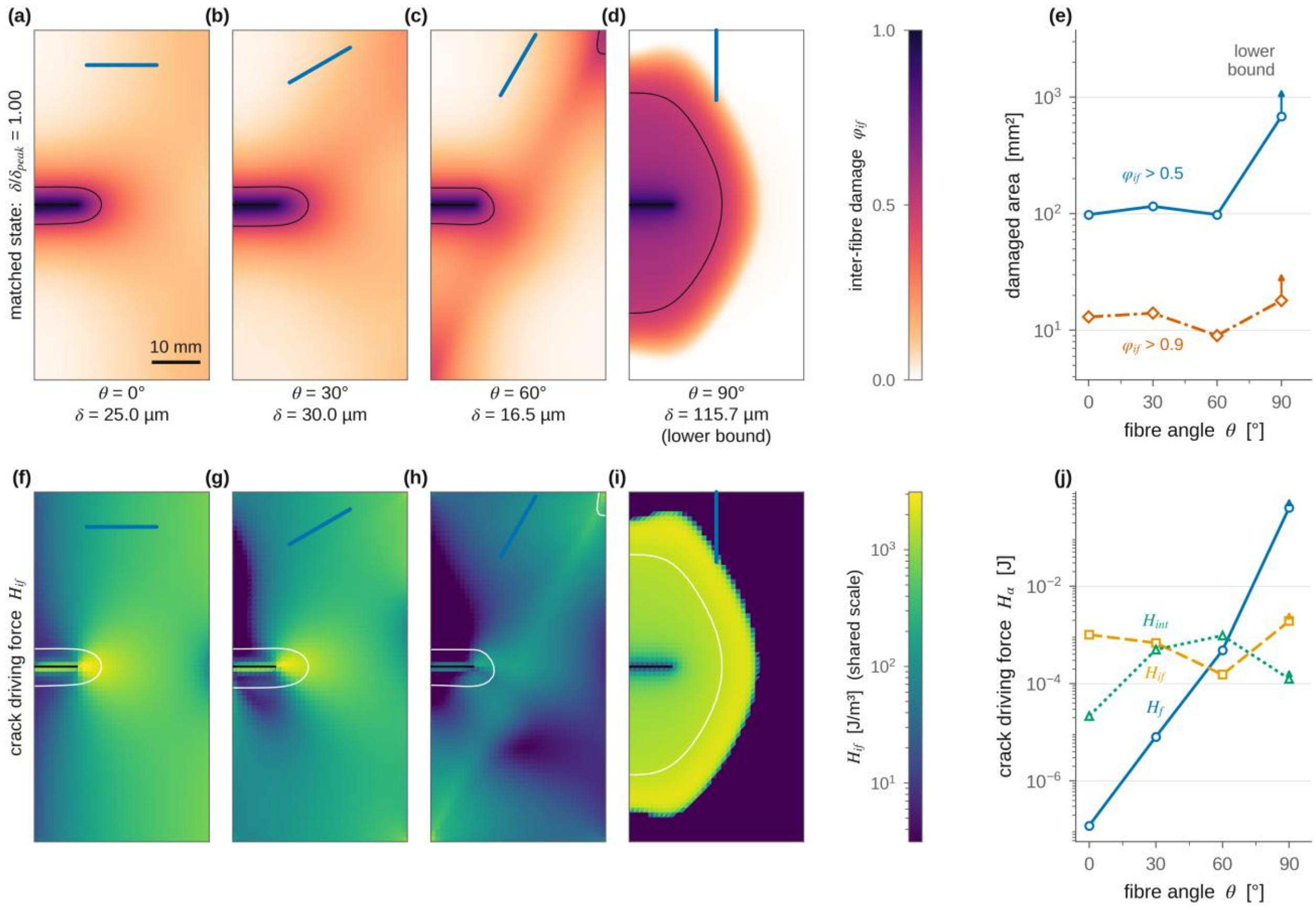

Fig. 12. Fibre orientation and the damage envelope at the matched state $\delta/\delta_{peak} = 1.00$. (a)–(d) Inter-fibre damage $\varphi_{if}$; material above $\varphi_{if} = 0.9$ is drawn as separated over the diffuse field and the blue segment gives the fibre direction. (e) Damaged area against fibre angle. (f)–(i) Inter-fibre crack driving force $H_{if}$ on a shared scale. (j) Specimen-integrated driving forces. The $\theta = 90°$ solve is arrested at a limit point, so its values are lower bounds, drawn with arrows.

## 7.12 Mechanism competition and load capacity

Fig. 13(a) follows the $\theta = 60°$ solve, which is the case with a fully resolved post-peak, with the damage field embedded at four stages. Stage (i) is the imposed notch seed of 85 nodes above $\varphi_{if} = 0.5$, present from $\delta = 0$ and unchanged to $\delta = 14.25$ µm, so the response to that point is elastic loading of a notched shell and only the growth beyond it is a computed result. By stage (iv) the band has left the notch line and elongated along the fibre direction, which is the post-peak deflection that Fig. 12 cannot show at the matched state because it has not yet developed there.

Panel (b) collects the four responses and panel (c) the peak reactions on a polar logarithmic scale, giving 82 N at 0°, 88 N at 30°, 194 N at 60° and at least 7933 N at 90°, a ratio of at least 96 over the sweep. The spread is not a stiffness effect, since the same operator, mesh and notch are used at every angle and the longitudinal-to-transverse modulus ratio is only 14; it follows instead from which mechanism has to carry the notch-tip field. Panel (d) makes this explicit as the share of the crack driving force carried by each mechanism at the matched state. At 0°, with the fibres perpendicular to the load, the inter-fibre channel carries essentially all of it. At 30° the interaction channel has taken about 40%, at 60° it dominates at about 60%, and at 90° the fibre channel carries the entire share. The transfer is monotone and it explains the capacity ordering, since the specimen is strongest when the mechanism carrying the driving force is the one with the highest toughness, and the toughness ratio $G_c^f/G_c = 1630$ is far larger than any stiffness ratio in the material. The sequence also identifies 60° as the angle at which two mechanisms are active together, which is why that case is used in Sections 7.7 and 7.13.

**Remark 8.** The fibre damage field never leaves its floor, with $\max\varphi_f = 3.0 \times 10^{-2}$ at every angle, so the dominant fibre share at 90° is stored elastically and is not dissipated. The driving force has moved to a channel that has not begun to fail, which is consistent with that solve being arrested at a limit point instead of softening. The share is defined at every angle only because it is computed from $H_\alpha$; of the signed contributions no share would exist at 90°, where the inter-fibre contribution is negative over 45% of the quadrature points.

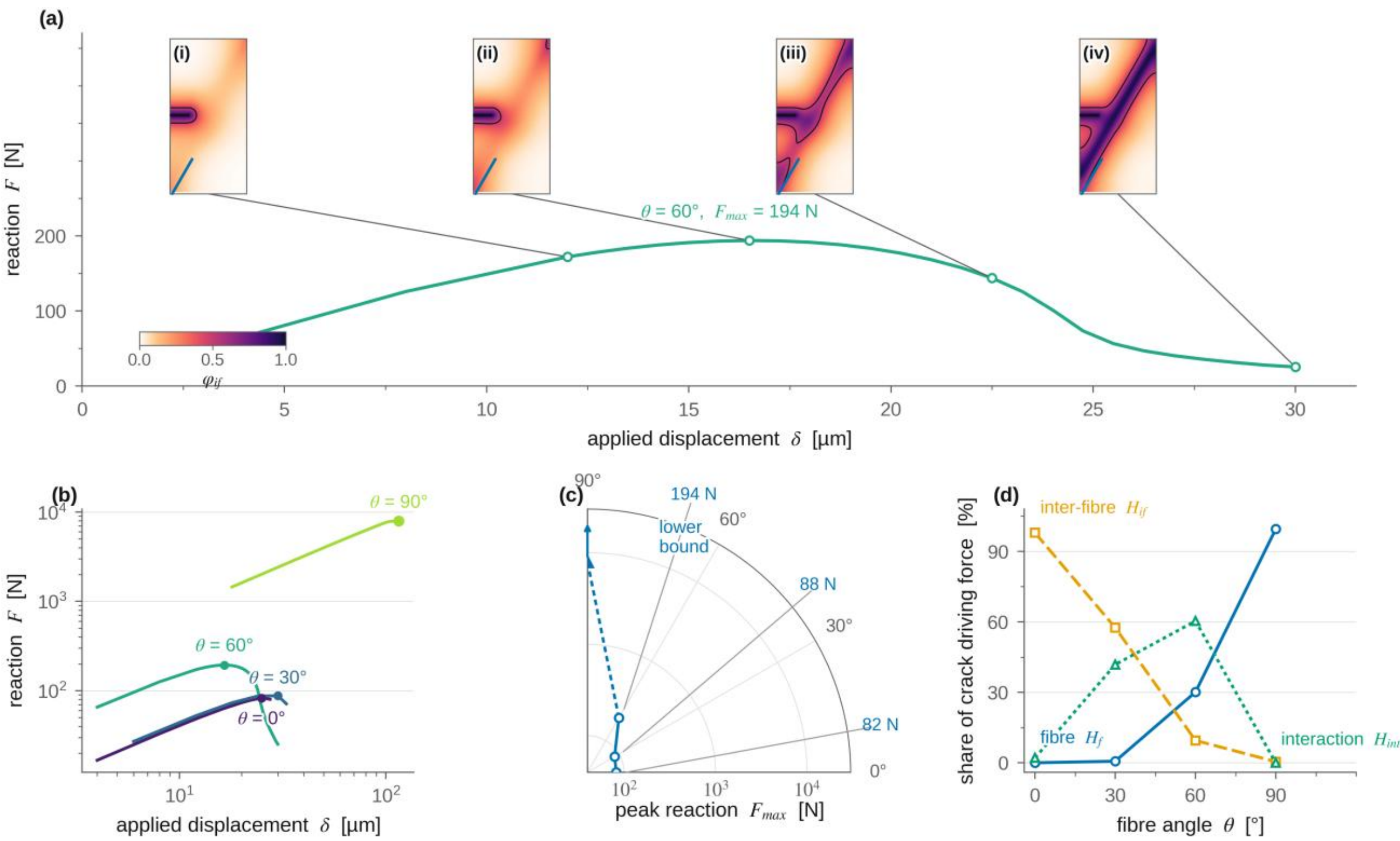

Fig. 13. Mechanism competition and load capacity. (a) Reaction against applied displacement at $\theta = 60°$ with the inter-fibre damage field at the four marked stages; stage (i) is the imposed notch seed. (b) Reaction for the four fibre-angle solves. (c) Peak reaction against fibre direction on a logarithmic radial scale. (d) Share of the crack driving force $H_\alpha$ carried by each mechanism at the matched state.

## 7.13 The interface field as a competing dissipation path

The interfacial field $\beta$ degrades only the fracture-active part of the interaction channel, so it couples to the mechanical fields through the strain rather than to the bulk damage fields, and it carries a driving term of its own in Eq. (25). Fig. 14 examines it at $\theta = 60°$, the angle at which the interaction channel carries most of the driving force. Panels (a) and (b) place the bulk damage and the interfacial field at the same state on one scale. They localise differently: the bulk band is notch-aligned while the interfacial band lies at 60.7° against a 60° fibre direction. The interfacial field therefore selects the fibre direction without any anisotropy in its own gradient term, since the direction is inherited from $J_{ft}$, which is built on the fibre projector of Eq. (7), rather than imposed on the crack surface density.

Panel (c) sweeps the interfacial toughness $G_c^{int}$ over 10, 50, 200 and 1000 J/m². After each field is normalised by its own maximum the four correlate at 0.994 or better, which separates the two controls: $G_c^{int}$ sets the amplitude of $\beta$ and $\ell_\beta$ sets where it localises. Panel (d) shows the consequence for the structural response against a $\beta$-off control on matched schedules. At 10 J/m² the peak falls from 193.8 N to 92.5 N, so the interfacial mechanism more than halves the capacity. At 200 J/m² the response is within 3.0% of the $\beta$-off twin and at 1000 J/m² within 0.7%, where $\beta$ is effectively inert. That last rung is within 3% of the interlaminar value of Camanho et al. [2], so for this specimen at that toughness the interface would not be the limiting mechanism.

Panel (e) resolves the crack-surface energy by channel and shows that the interface does not simply add dissipation but takes it over. At the last state the interface carries 0.82 mJ while the bulk crack is held to 0.76 mJ, against 3.76 mJ of bulk dissipation in the run without the interfacial field. The bulk crack is arrested because the interface has removed the driving energy that would have advanced it, which is the mechanism competition of Section 7.12 expressed in a different variable.

**Remark 9.** Two limitations attach to this study. The interfacial toughness is not calibrated, and the sweep is reported as the evidence with no value fitted. In addition, $\ell_\beta = 5$ mm is coarse for a 36 mm specimen: the measured band is 17 mm wide, about 1.7 times the width its own regularisation length permits, so the band sits at its resolution floor and no toughness in this sweep produces a narrow one. The amplitude conclusions are unaffected, but the width of the interfacial band should not be read as a physical length.

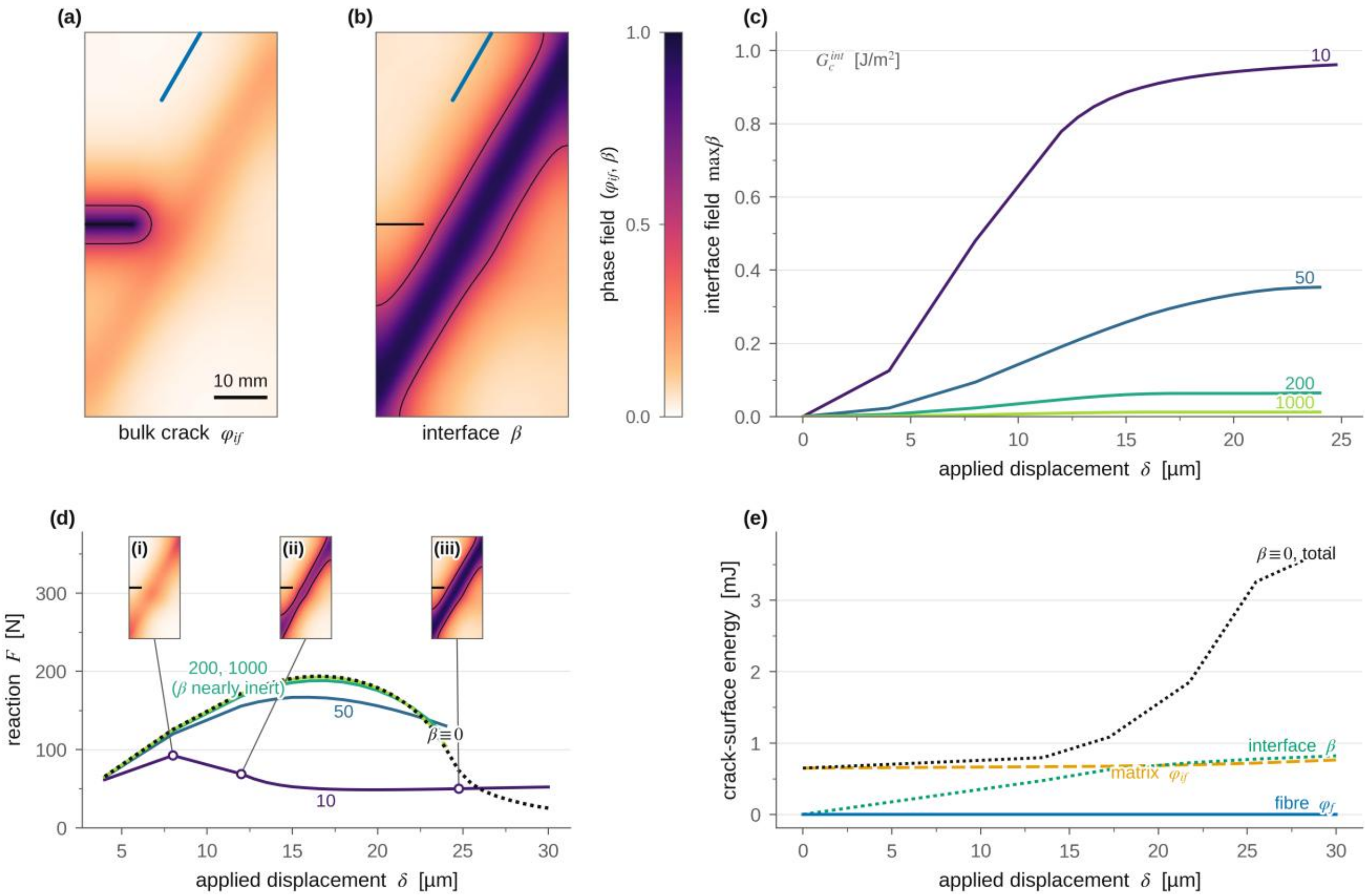

Fig. 14. The interfacial field at a fibre angle of 60°. (a, b) Bulk inter-fibre damage and the interfacial field at the same state on one shared scale. (c) Peak interfacial field over the toughness sweep. (d) Load–displacement with and without the interfacial term on matched schedules. (e) Crack-surface energy by channel for the coupled run against the total of the control. The control retains the interaction term of Eq. (14) and freezes $\beta$ at zero, so it differs from the coupled runs in one variable only; a control run under a different degradation of the interaction channel would differ in two.

### 7.14 Curvature-induced through-thickness asymmetry

The surface-resolved parameterisation of Section 3.3 carries two independent damage fields on the two faces of the shell, and the question it raises is when that capability is required rather than merely available. Fig. 15 answers it with curvature. On a flat panel the strain is symmetric about the midsurface under in-plane loading and the two fields coincide; on a curved panel the shifter of Eq. (2) makes the invariants vary through the thickness, as measured in Fig. 5(c), and the two faces are no longer driven equally.

Panels (a) and (b) show the upper and lower surface fields on the curved strip at $R = 25$ mm under circumferential loading, at the last converged state of that solve, on one shared damage scale. The seeded notch is common to both surfaces and present from zero load, so only the growth beyond it is a result; the lower surface carries the longer crack. Panel (c) shows their difference on a diverging scale with the zero contour drawn. The asymmetry is one-signed, attains $d_u - d_l = -0.429$ at its extremum and is localised ahead of the notch tip, which is where the through-thickness strain gradient is largest.

To quantify the membrane–bending state of the curved strip, the invariant norms of the membrane and bending resultants were evaluated at a common elastic load ($\delta \approx 10\mu$m). For circumferential loading, $6\|\mathbf{M}\|/(h\|\mathbf{N}\|)$ increases from 7.00 at $R = 100$ mm to 10.64 at $R = 50$ mm and 15.46 at $R = 25$ mm, indicating an increasingly bending-dominated response as curvature increases. At $R = 25$ mm, for example, $\|\mathbf{N}\|_{max} = 1.24 \times 10^2$ N/m and $\|\mathbf{M}\|_{max} = 0.168$ N. Under axial loading, the corresponding ratios are 0.62, 0.36 and 0.32, indicating a membrane-dominated response. In the flat limit, $\|\mathbf{M}\|$ is at machine precision, confirming that the bending resultant is induced by curvature.

Panel (d) is the control. The same difference for the flat pair is not merely small at a matched state but identically zero at every one of that solve's thirty-one steps, at most $6 \times 10^{-16}$. A flat panel at this fibre angle carries no through-thickness asymmetry, so the field plotted in panel (c) measures curvature alone and no residual asymmetry of numerical origin can be mistaken for the effect.

Panel (e) collects the separation between the two loading configurations, $\Delta = \| d_u - d_l \|_{circ} - \| d_u - d_l \|_{axial}$, against inverse radius at four radii, each read at one common damage level rather than at each solve's own end point, since the legs terminate at different states. The flat value is $-1.9 \times 10^{-14}$, so the curved values stand about thirteen orders above it. The separation is monotone in $1/R$ and follows $\Delta \sim (1/R)^{0.375}$. It is therefore sub-linear, and the best line through the origin departs from the data by 51 per cent. An $O(h/R)$ argument applied to the strain asymmetry alone would predict a linear law, so damage does not inherit the scaling of the field that drives it, which is consistent with a threshold process in which the driving force must exceed a critical value before any of it is converted into crack surface. The exponent is reported at one mesh and one reading level, both stated in the caption.

Two limitations attach to this result. The thin limit is not shown: pairing curvature with thickness would require a surface-resolved thickness ladder, whereas the thickness legs available here are mechanism-resolved, with damage uniform through the thickness and no route for the observable to vanish in $h$. And the study demonstrates a formulation capability rather than a calibrated structural prediction, since the interfacial toughness is uncalibrated and the geometry is a benchmark.

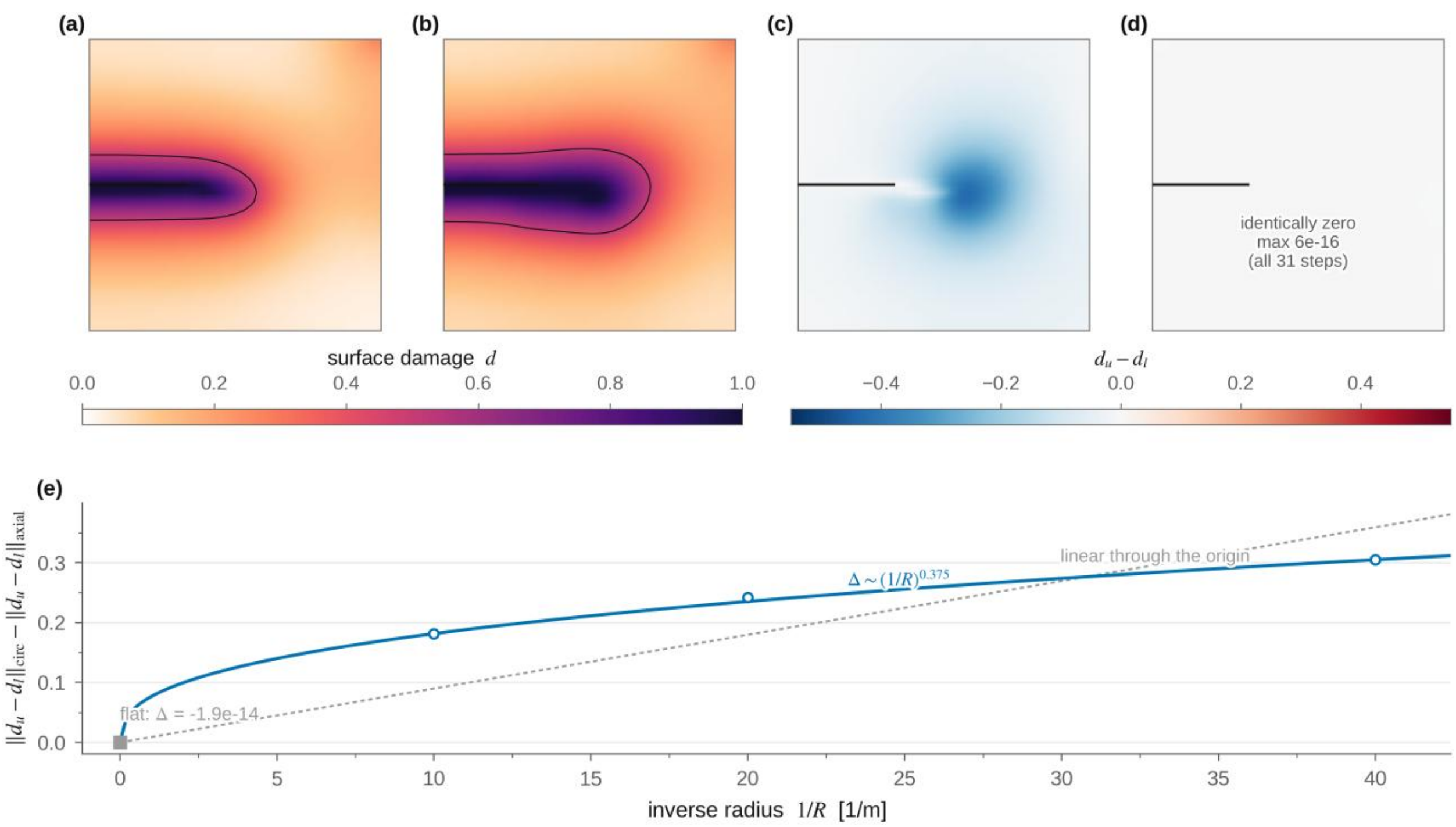

Fig. 15. Surface-resolved damage on the curved strip (IM7/8552, fibre at 45°, $36 \times 36$ elements). (a, b) Upper and lower surface fields at $R = 25$ mm under circumferential loading at the last converged state, on one shared scale fixed to 0–1; the hatched segment is the seeded notch, common to both surfaces. (c) Their difference on a diverging scale with the zero contour drawn. (d) The same difference for the flat pair on the same scale,

identically zero at every step. (e) Separation $\Delta$ between the two loading configurations against inverse radius at four radii, read at one common damage level; the dashed reference is the best line through the origin.

### 7.15 Summary of the numerical studies

The studies establish four points. The element does not lock on an exact inextensional field over a sixteen-fold thickness range, and on curved geometry the chain recovers the flat solution as the radius grows while the invariants remain independent of the chart. The discrete operators reproduce the first and second variations of a single potential in every field block, the plane-stress condensation and the crack functional behave as derived, and the assembled tangent is definite along the equilibrium path up to the deep end of softening. The certified model then resolves mechanism-specific behaviour: fibre orientation governs the damage envelope and the load capacity, the interfacial field opens a competing fibre-aligned dissipation path, and curvature drives a through-thickness asymmetry that two surface fields resolve and one cannot. A learned operator seated in the same socket is admissible and reproduces the closed form at interpolation scale, but is not deployed in any result reported here. Each point is measured against a control able to fail, and the limitations of individual studies are stated in the accompanying remarks.

## 8. Conclusions

This work has developed a phase-field fracture formulation for transversely isotropic composite shells in which the constitutive law is a replaceable operator rather than a fixed expression. The main conclusions are as follows.

First, the Cholesky pull-back of the Green–Lagrange strain onto an orthogonal invariant basis yields a closure identity that is exact, so the separation of the stored energy into fibre, inter-fibre and interaction channels partitions the strain norm without residue and introduces no modelling freedom beyond the choice of the three channel potentials. Because the residual, the tangent, the plane-stress condensation and the assembly are written in those potentials and their derivatives alone, one solver serves a closed-form and a learned representation without change. The two properties most at risk in that exchange are proved to survive it: the geometric tangent is independent of the material tangent, and the condensation remains regular wherever the residual stiffness is positive.

Second, the numerical evidence supports each of these statements against a control capable of failing. The second variation converges at $1.4 \times 10^{-11}$ with the geometric term and stalls at $1.4 \times 10^{-5}$ without it, a difference that quadratic convergence of the Newton iteration does not detect. The condensation holds the through-thickness stress about seventeen decades below the uncondensed control, and the derivative governing its regularity stays above 3.244 GPa through damage to $d = 0.77$, so the residual stiffness introduced for numerical safety is also what keeps the condensation well posed. Counted exactly by pivot signs, the assembled tangent is positive semi-definite through the peak and to 14% of the peak load on the descending branch, losing definiteness only at the deep end of softening with three and then four negative modes; reporting the count rather than a verdict is what makes this signal usable. The crack functional recovers the prescribed toughness to within 0.7%, with the residual error governed by the resolution ratio alone, the peak load is mesh-objective to 0.20% over the two finest rungs, and a published anisotropic energy maps into the present invariant basis at $3.4 \times 10^{-16}$, thirteen decades below a deliberately wrong mapping.

Third, the discretisation is verified before it is used. On an exact inextensional deformation the deployed element follows $\Pi \sim t^{2.937}$ against the exact cubic bending law over a sixteen-fold thickness range, while the same element with a planted rotation-frame defect follows $\Pi \sim t^{1.037}$; under mesh refinement the spread of $\Pi/t^3$ across the ladder moves towards the exact law for the deployed element and away from it for the defect, which is the signature that separates locking from discretisation error. No assumed-strain treatment is therefore required in the range examined. On curved geometry the chain recovers the flat solution at order 1.72 in $1/R$, and the four invariants agree across three charts of the same surface to $2.5 \times 10^{-15}$ while chart-dependent components of the same states differ by order unity, fifteen decades above.

Fourth, the mechanism channels are signed invariant-group contributions rather than energies of separable subsystems, so shielding under transverse contraction is realised as an inactive damage constraint and is reported through the pointwise positive part of each channel, at a measured additivity cost of $2.0 \times 10^{-2}$ at 90° and zero elsewhere. On this basis, fibre orientation controls the damage envelope and the load capacity but not the crack path: the envelope area varies by a factor of 7.0 over 0–90° and the peak reaction by at least a factor of 96, while the separated material stays at the notch. The ordering follows the transfer of driving force from the inter-fibre channel at 0° through the interaction channel at 60° to the fibre channel at 90°, and is set by the toughness ratio $G_c^f/G_c = 1630$ rather than by any stiffness contrast. The interfacial field behaves as a distinct mechanism rather than a correction, localising along the fibre direction while the bulk band stays notch-aligned and, at low interfacial toughness, taking dissipation over from the bulk crack and halving the peak capacity.

Fifth, curvature drives a through-thickness fracture asymmetry that two surface fields resolve and one midsurface field cannot. The difference between the surface fields is identically zero on the flat panel at every step of that solve, at most $6 \times 10^{-16}$, and reaches $|d_u - d_l| = 0.429$ at $R = 25$ mm. The separation between the two loading configurations grows monotonically with inverse radius as $(1/R)^{0.375}$ rather than by the linear law that an $O(h/R)$ argument applied to the strain asymmetry alone would predict, departing from the best line through the origin by 51 per cent; damage does not inherit the scaling of the field that drives it. The flat control is what makes this reading available, since it removes any residual asymmetry of numerical origin from the comparison.

Finally, the framework separates the question of admissibility from the question of prediction. The admissibility conditions state what a candidate operator must satisfy, the three tiers state what must be checked before it is used, and the corpus admitted under Tier C is by construction a set of internally consistent equilibrium states of the model being solved. Predictive learned constitutive modelling on this socket can therefore be posed as a question about the operator alone. Three items remain open and are separable from the framework itself: convexity of the two quadratic invariants in the deformation gradient, which is currently discharged by a pointwise certificate rather than proved; calibration of the interfacial toughness together with a regularisation length fine enough to resolve the interfacial band; and experimental validation of the orientation-dependent capacity, for which the envelope area, the peak reaction and the per-mechanism driving share reported here are the natural comparison targets.

## CRediT authorship contribution statement

**Altaf Ahmad Lone:** Writing – original draft, Visualization, Validation, Software, Methodology, Investigation, Formal analysis, Data curation, Conceptualization. **R. T. Durai Prabhakaran:** Writing – review & editing, Supervision, Resources, Project administration. **Timon Rabczuk:** Writing – review & editing, Supervision, Methodology, Conceptualization. **Xiaoying Zhuang:** Writing – review & editing, Supervision, Methodology.

## Declaration of competing interest

The authors declare that they have no known competing financial interests or personal relationships that could have appeared to influence the work reported in this paper.

## Acknowledgements

The first author gratefully acknowledges the support of the Indian Institute of Technology Jammu. The authors thank the Department of Mechanical Engineering, IIT Jammu, for access to computational resources.

## Appendix A. Admissible constitutive operators: definitions, proofs and equilibrium-based learning

An admissible $\Psi_\alpha$ (Definition A.1) determines $\Psi_d$ through Eq. (13) or Eq. (14) and hence $\Pi$ through Eq. (18); differentiating $\Pi$ once and twice gives the residual and tangent of Section 4, and discretising and solving that system as in Section 5 produces the states certified by Tiers A and B of Section 6. What each step guarantees differs in kind. The variational structure follows from calculus alone: any $\Psi_\alpha$ satisfying (A1)–(A6) yields a well-defined, symmetric $\mathbf{K}$ with $\mathbf{K}_{geo}$ present regardless of representation (Appendix A.6). Two admissibility properties are proved outright, namely convexity of $I_f$ and $I_t$ directly in $\mathbf{F}$ (Appendix A.3) and branchwise coercivity at the reference state (Appendix A.7); two are left open with a local numerical certificate in their place, namely polyconvexity of $J_t$ and $J_{ft}$ (Appendix A.4) and Newton convergence outside a kink-safe neighbourhood (Appendix A.8). Tier C is a definition rather than a theorem: it certifies that every state used to shape a learned $\Psi_\alpha$ is an internally consistent equilibrium of the model being solved, not that the resulting $\Psi_\alpha$ recovers the true material law.

### A.1 Definition (admissible constitutive operator)

A mapping $\Psi_\alpha: \mathcal{J}_\alpha \to \mathbb{R}$, $\alpha \in \{f, mx, int\}$, is admissible for Eq. (10) if it satisfies the following six conditions.

**(A1) Regularity.** $\Psi_\alpha \in C^k(\mathcal{J}_\alpha)$, with $k \geq 2$ for $\mathbb{C}$ to exist (Eq. (15)) and $k \geq 3$ for a Lipschitz tangent and a quantitative Newton convergence radius (Appendix A.8).

**(A2) Objectivity.** Automatic: $\mathcal{J}_\alpha$ is built from $\hat{\mathbf{E}} = \mathbf{L}^{-1}\mathbf{E}\mathbf{L}^{-\mathsf{T}}$ (Eq. (6)), and $\mathbf{E}$ is already objective (Eq. (5)).

**(A3) Stress-free reference.** $\Psi_\alpha(\mathbf{0}) = 0$ and $D\Psi_\alpha(\mathbf{0}) = \mathbf{0}$ at the origin of $\mathcal{J}_\alpha$, since $\mathbf{E} = \mathbf{0} \Rightarrow \hat{\mathbf{E}} = \mathbf{0} \Rightarrow \mathcal{J}_\alpha = \mathbf{0}$ for these strain-built invariants. For a raw network $\widetilde{\Psi}_\alpha$ carrying no such guarantee, use

$$\Psi_\alpha(\mathcal{J}_\alpha) := \widetilde{\Psi}_\alpha(\mathcal{J}_\alpha) - \widetilde{\Psi}_\alpha(\mathbf{0}) - D\widetilde{\Psi}_\alpha(\mathbf{0}) \cdot \mathcal{J}_\alpha \qquad (\text{A}.1)$$

which preserves convexity, since subtracting an affine function of the argument does not change a Hessian.

**(A4) Convexity, per invariant.** From Eq. (8), $I_f$ and $I_t$ are affine in $\hat{\mathbf{E}}$, while $J_t$ and $J_{ft}$ are positive-semidefinite quadratic, hence convex, in $\hat{\mathbf{E}}$ (Appendix A.2). Consequently $\Psi_\alpha$ convex in $I_f$ and $I_t$ is sufficient with no monotonicity condition, while $\Psi_\alpha$ convex and non-decreasing in $J_t$ and $J_{ft}$ is required: composing a convex function with an affine inner map preserves convexity regardless of monotonicity, whereas composing with a convex non-affine inner map preserves it only if the outer function is monotone in the right direction. This gives convexity of $\Psi_\alpha\left(\mathcal{J}_\alpha(\hat{\mathbf{E}})\right)$ in $\hat{\mathbf{E}}$, hence in $\mathbf{E}$ through the linear congruence of Eq. (6). Whether it gives convexity in $\mathbf{F}$ is addressed in Appendices A.3 and A.4: for $I_f$ and $I_t$ the answer is yes; for $J_t$ and $J_{ft}$ it remains open.

An input-convex network [38] supplies (A4) on the $J_t$ and $J_{ft}$ slots through non-negative weights and a convex, non-decreasing activation, but that constraint is unnecessarily strong on the $I_f$ and $I_t$ slots, where unconstrained weights already preserve convexity in $\hat{\mathbf{E}}$. One convex network per mechanism, with non-negativity applied only to the $(J_t, J_{ft})$ channels, is therefore both correct and less restrictive.

**(A5) Growth and coercivity.** Independent of (A4) [55]: $\Psi_\alpha \to \infty$ as admissible deformations approach the boundary of physical admissibility. This is not implied by convexity, and not automatic for a saturating-activation network without an explicit unbounded term.

**(A6) Material symmetry.** Discharged by construction: the orthogonal identity of Eq. (9) is the transversely isotropic Spencer–Boehler structural-tensor basis [57–59], specialised to phase-field form following Teichtmeister et al. [26]. Any $\Psi_\alpha$ built on $\mathcal{J}$ respects transverse isotropy about $\hat{\mathbf{m}}$ automatically, conditional on $\hat{\mathbf{m}}$ being the material's symmetry axis, which is an assumption on input data rather than a provable property of the mathematics (Appendix A.9).

### A.2 Convexity of $J_t$ and $J_{ft}$ in $\hat{\mathbf{E}}$

**Claim.** $J_{ft}(\hat{\mathbf{E}}) = 2 \parallel \mathbf{P}_f \hat{\mathbf{E}} \mathbf{P}_t \parallel^2$ is a positive-semidefinite quadratic form in $\hat{\mathbf{E}}$, hence convex on all of $\mathrm{Sym}(3)$; likewise $J_t(\hat{\mathbf{E}}) = \parallel \mathrm{dev}\mathbf{E}_t \parallel^2$.

*Proof.* The projectors $\mathbf{P}_f$ and $\mathbf{P}_t$ of Eq. (7) are independent of $\hat{\mathbf{E}}$, so $\mathbb{T}(\hat{\mathbf{E}}) := \mathbf{P}_f \hat{\mathbf{E}} \mathbf{P}_t$ is linear in $\hat{\mathbf{E}}$. Then $J_{ft} = 2\,\mathbb{T}(\hat{\mathbf{E}}) : \mathbb{T}(\hat{\mathbf{E}}) = \hat{\mathbf{E}} : \mathbb{A} : \hat{\mathbf{E}}$ with $\mathbb{A} = 2\,\mathbb{T}^*\mathbb{T} \succcurlyeq \mathbf{0}$, a positive-semidefinite quadratic form and therefore convex everywhere. The identical argument applies to $J_t$ with $\mathbb{T}(\hat{\mathbf{E}}) = \mathrm{dev}(\mathbf{P}_t \hat{\mathbf{E}} \mathbf{P}_t)$. ■

This also gives a second proof that every term in Eq. (9) is nonnegative, which is why $J_t$ and $J_{ft}$ are left unsplit by the Macaulay bracket in Eq. (11): there is no sign to split on. An equivalent alternative for $\Psi_{int}$, with $I_5 := \parallel \hat{\mathbf{E}}\hat{\mathbf{m}} \parallel^2$ and $I_4 := I_f$, is

$$I_5 - I_4^2 = \parallel (\mathbf{I} - \hat{\mathbf{m}} \otimes \hat{\mathbf{m}})\hat{\mathbf{E}}\hat{\mathbf{m}} \parallel^2 \qquad (\text{A}.2)$$

obtained by decomposing $\hat{\mathbf{E}}\hat{\mathbf{m}}$ into components parallel and perpendicular to $\hat{\mathbf{m}}$. It has the same positive-semidefinite quadratic structure as $J_{ft}$ and is an equally valid choice for (A4).

### A.3 Convexity of $I_f$ and $I_t$ directly in F

Convexity in $\hat{\mathbf{E}}$ transfers to convexity in $\mathbf{E}$ through the linear congruence of Eq. (6), but $\mathbf{E} = 1/2\,(\mathbf{F}^{\mathsf{T}}\mathbf{F} - \mathbf{G}_0)$ is quadratic in $\mathbf{F}$, and convexity does not automatically survive a non-affine reparameterisation. The following closes this gap for the two linear invariants.

**Lemma.** Let $\mathbf{Q}$ be any fixed, symmetric, positive-semidefinite tensor built only from reference data, and let $q(\hat{\mathbf{E}}) := \hat{\mathbf{E}}\!:\!\mathbf{Q}$. Pulling back through Eq. (6) gives $q = \mathbf{E}\!:\!\mathbf{N}$ for the fixed positive-semidefinite tensor $\mathbf{N} := \mathbf{L}^{-\mathsf{T}}\mathbf{Q}\mathbf{L}^{-1}$, since congruence preserves positive-semidefiniteness. Writing $\mathbf{N} = \sum_{i=1}^{r}\mu_i\ \mathbf{w}_i \otimes \mathbf{w}_i$ with $\mu_i > 0$ and using $\mathbf{E} = 1/2\,(\mathbf{F}^{\mathsf{T}}\mathbf{F} - \mathbf{G}_0)$,

$$q(\mathbf{F}) = \frac{1}{2}\sum_{i=1}^{r}\mu_i\left(\parallel \mathbf{F}\mathbf{w}_i \parallel^2 - \mathbf{w}_i^{\mathsf{T}}\mathbf{G}_0\mathbf{w}_i\right) \qquad (\mathrm{A.3})$$

a nonnegative combination of squared norms of fixed linear maps of $\mathbf{F}$, each convex because $\mathbf{F} \mapsto \mathbf{F}\mathbf{w}_i$ is linear, shifted by a constant. Hence $q$ is convex in $\mathbf{F}$ directly, not merely polyconvex and not merely convex in $\mathbf{E}$. ■

**Corollary.** $I_f$ and $I_t$ both satisfy the hypothesis of the Lemma and are therefore both convex in $\mathbf{F}$ directly. For $I_f$, take $\mathbf{Q} = \mathbf{P}_f = \hat{\mathbf{m}} \otimes \hat{\mathbf{m}}$, of rank one. Since $\hat{\mathbf{m}} = \mathbf{L}^{\mathsf{T}}\mathbf{m}$, the pull-back collapses exactly, $\mathbf{L}^{-\mathsf{T}}\mathbf{P}_f\mathbf{L}^{-1} = \mathbf{m} \otimes \mathbf{m}$, giving the single-term case of Eq. (A.3),

$$I_f(\mathbf{F}) = \frac{1}{2}(\parallel \mathbf{F}\mathbf{m} \parallel^2 - 1) \qquad (\mathrm{A.4})$$

using $\mathbf{m}^{\mathsf{T}}\mathbf{G}_0\mathbf{m} = 1$. This is the fibre-stretch invariant of finite-strain anisotropic elasticity, on which the Holzapfel–Gasser–Ogden $\langle I_4 - 1\rangle$ family builds for $\mathbf{C}$-based invariants; the construction of Eq. (6) inherits it unchanged. For $I_t$, take $\mathbf{Q} = \mathbf{P}_t = \mathbf{I} - \mathbf{P}_f$, of rank two. The pull-back gives $\mathbf{N}_t = \mathbf{G}_0^{-1} - \mathbf{m} \otimes \mathbf{m}$, which is positive-semidefinite: writing $\mathbf{x} := \mathbf{G}_0^{1/2}\mathbf{m}$, the normalisation $\parallel \mathbf{x} \parallel^2 = \mathbf{m}^{\mathsf{T}}\mathbf{G}_0\mathbf{m} = 1$ gives $\mathbf{I} - \mathbf{x} \otimes \mathbf{x} \succcurlyeq \mathbf{0}$, and transforming congruently by $\mathbf{G}_0^{-1/2}$ with $\mathbf{G}_0^{-1/2}\mathbf{x} = \mathbf{m}$ gives $\mathbf{G}_0^{-1} - \mathbf{m} \otimes \mathbf{m} \succcurlyeq \mathbf{0}$ directly, of rank two with kernel along $\mathbf{G}_0\mathbf{m}$. The Lemma therefore applies and $I_t$ is convex in $\mathbf{F}$ directly.

### A.4 What remains open for $J_t$ and $J_{ft}$

The Lemma of Appendix A.3 does not reach $J_t$ and $J_{ft}$: they are quadratic rather than linear in $\hat{\mathbf{E}}$, hence quartic in $\mathbf{F}$, and a nonnegative combination of squared linear maps of $\mathbf{F}$ produces only a quadratic polynomial in $\mathbf{F}$, never a quartic one. Two routes close this, neither automatic. The structural route [42,52,53] builds the transverse-isotropy invariant directly from $\mathbf{F}$, as $I_1^{ti} := \mathrm{tr}(\mathbf{C}\mathbf{M}) = \parallel \mathbf{F}\mathbf{M} \parallel^2$, which is trivially convex in $\mathbf{F}$; closing $J_t$ and $J_{ft}$ this way requires re-deriving them as squared norms of linear maps of $\mathbf{F}$, carrying the Cholesky pull-back of Eq. (6) into the structural tensor rather than through $\mathbf{E}$. The local-certificate route [40] instead checks positive-definiteness of the relevant Hessian blocks numerically at each certification point visited during training or validation, which is the role of Tier A in Section 6. The local-certificate route is adopted here; the structural route remains open.

### A.5 Regularity of the plane-stress condensation

$\mathbf{S} = \partial\Psi_d/\partial\mathbf{E}$ is $C^{k-1}$ if $\Psi_d \in C^k$. Solving $S_{33}(E_{33}; E_{\alpha\beta}) = 0$ for $E_{33}^*(E_{\alpha\beta})$ by the implicit function theorem preserves $C^{k-1}$ regularity of $E_{33}^*$ whenever $\partial S_{33}/\partial E_{33} = \mathbb{C}_{3333} \neq 0$. Since $g(d) \geq k > 0$ throughout (Eq. (13)), the fracture-active part of $\mathbb{C}_{3333}$ is scaled by $g(d)$ and remains bounded below by $k$ times its undamaged value, so $\mathbb{C}_{3333} > 0$ even in the limit $d \to 1$. The residual stiffness already present for numerical safety is therefore exactly what keeps the condensation of Eq. (16) non-degenerate.

### A.6 The structural tangent: $\mathbf{K}_{geo} \neq \mathbf{0}$

Write $\mathbf{w} = (\mathbf{u}, \boldsymbol{\theta})$. From Eq. (3), $\mathbf{F}(\mathbf{w}) = \mathbf{F}_0 + \mathbb{L}[\mathbf{w}]$ is affine in $\mathbf{w}$, the linearised-director choice keeping $D^2\mathbf{F} = \mathbf{0}$ exactly where an exact-rotation director would not. With $\mathbf{E} = 1/2\,(\mathbf{F}^{\mathsf{T}}\mathbf{F} - \mathbf{G}_0)$,

$$D^2\mathbf{E}[\Delta\mathbf{w}, \Delta\mathbf{w}] = (\mathbb{L}[\Delta\mathbf{w}])^{\mathsf{T}}(\mathbb{L}[\Delta\mathbf{w}]) \succcurlyeq \mathbf{0} \qquad (\mathrm{A.5})$$

which is generically nonzero. *Proof.* Differentiate $\mathbf{E}(\mathbf{w})$ twice. The first derivative $D\mathbf{F}[\Delta\mathbf{w}] = \mathbb{L}[\Delta\mathbf{w}]$ is constant in $\mathbf{w}$, and $D^2\mathbf{F} = \mathbf{0}$ removes the two product-rule terms that would otherwise involve it, leaving the Gram term above. ■ Hence the geometric tangent is the bilinear form

$$\mathbf{K}_{geo}[\delta\mathbf{w}, \Delta\mathbf{w}] = \int_{\Omega_0} \mathbf{S} : \left[(\mathbb{L}[\delta\mathbf{w}])^{\mathsf{T}}\mathbb{L}[\Delta\mathbf{w}]\right] \mathrm{d}V \qquad (\mathrm{A.6})$$

which depends on $\mathbf{S}$ and the kinematic Gram operator $\mathbb{L}^{\mathsf{T}}\mathbb{L}$, never on $\mathbb{C}$. Exchanging a learned $\Psi_\alpha$ for a closed-form one rescales $\mathbf{K}_{mat}$ but never removes $\mathbf{K}_{geo}$, so Eq. (28) must keep both terms regardless of how $\Psi_\alpha$ is represented.

### A.7 The Macaulay kink and branchwise coercivity

The Macaulay split of Eq. (11) composes each mechanism potential with the ramp $\langle\cdot\rangle_+ = \max(\cdot\,,0)$ applied to the linear switching functionals $I_f$ and $I_t$. For a general admissible $\Psi_\alpha$ this is only $C^0$ at a switching point $q = 0$: the ramp has a Heaviside-discontinuous derivative, and composing it with a smooth $\Psi_\alpha$ leaves a jump of size $D\Psi_\alpha(0)$ unless that derivative vanishes there. No choice of activation function inside $\Psi_\alpha$ removes this, since the split sits downstream of $\Psi_\alpha$ in the operator chain of Eqs. (6)–(15). Consequently $\Psi_d$, and hence $\mathbb{C}$, is at best piecewise $C^k$ with genuine kinks on $\{I_f = 0\} \cup \{I_t = 0\}$, and any Newton-convergence argument built on $\delta^3\Pi$ is valid only on a kink-safe region bounded away from these surfaces.

At the reference state the picture improves, because (A3) supplies the missing condition: $D\Psi_\alpha(\mathbf{0}) = \mathbf{0}$ removes the jump at $\hat{\mathbf{E}} = \mathbf{0}$. Writing $\eta(x) = \langle x\rangle_+^2$, which is $C^1(\mathbb{R})$ with $\eta(0) = \eta'(0) = 0$, convex and non-decreasing, and with only $\eta''$ discontinuous at $x = 0$, a Taylor expansion of $\Psi_\alpha$ about the origin shows that the switch terms behave to leading order like a multiple of $\eta\left(q(\hat{\mathbf{E}})\right)$ for each linear switching functional $q \in \{I_f, I_t\}$, which is positive-semidefinite quadratic near $\hat{\mathbf{E}} = \mathbf{0}$ by the composition rule of (A4). This local structure is what the coercivity estimate below relies on.

**Lemma.** Write $\Psi_d = \Psi_{smooth} + \Psi_{kink}$, collecting the switch terms in $\Psi_{kink}$, each nonnegative and vanishing to first order at $\hat{\mathbf{E}} = \mathbf{0}$ by the argument above, with $\Psi_{smooth}$ genuinely $C^3$. If $D^2\Psi_{smooth}(\mathbf{0}) \succcurlyeq c_0\mathbf{I}$ with $c_0 > 0$, then for $\parallel \mathbf{H} \parallel$ small,

$$\Psi_d(\mathbf{H}) - \Psi_d(\mathbf{0}) - D\Psi_d(\mathbf{0}) \cdot \mathbf{H} \;\geq\; \frac{c_0}{2} \parallel \mathbf{H} \parallel^2 - C \parallel \mathbf{H} \parallel^3 \;\geq\; \frac{c_0}{4} \parallel \mathbf{H} \parallel^2 \qquad (\mathrm{A.7})$$

using the $C^3$ Taylor remainder on $\Psi_{smooth}$ and $\Psi_{kink} \geq 0$ to drop the second bracket, despite $\Psi_d \notin C^2$ at $\hat{\mathbf{E}} = \mathbf{0}$. This removes the local coercivity obstruction near the reference state. It does not remove the harder step of constructing equilibria by the implicit function theorem, which requires two-sided $C^1$ regularity and fails exactly at the switch; that step is closed only by semismooth, Clarke-Jacobian machinery.

### A.8 Newton–Kantorovich convergence

Newton's method, Eq. (31), converges quadratically under the Newton–Kantorovich theorem [60–62] provided $\mathbf{K} = D^2\Pi$ is Lipschitz near the solution, which follows from $\Pi \in C^3$. Given Appendix A.7, this can be claimed only on a kink-safe subset, and only if $\Psi_\alpha$ is built from $C^3$ activations, which rules out the ReLU and LeakyReLU families and favours Softplus or a spline construction. The residual stiffness $k$ in Eq. (13) does double duty: it is the usual safeguard against a singular tangent at full damage, and, by Appendix A.5, it is also what keeps the plane-stress condensation well posed there.

### A.9 Equilibrium-based training of admissible operators

Definition A.1 states what $\Psi_\alpha$ must satisfy, not how it is obtained. Li and Chen [44] train a network to output stress as an internal variable, supervised not by stress labels but by requiring the predicted field to satisfy nodal-force equilibrium, which is the residual structure of Eqs. (23)–(26) used as a training loss rather than as a Newton residual. Li et al. [45] extend this to history-dependent elastoplasticity by an incremental update with three sub-networks — an input-convex network for the equivalent stress and two further networks for yield strength and internal-variable evolution — with no assumed yield function or flow rule. Li et al. [46] extend the same paradigm to three-dimensional anisotropic hyperelasticity: an input-convex network over invariant-formed inputs, with local polyconvexity enforced by a Hessian-eigenvalue penalty, the direct analogue of Tier A in Section 6, followed by a symmetry-identification step that searches for the rotation group under which the trained stress response is invariant and fine-tunes to respect it by Reynolds-operator averaging. A related but distinct paradigm minimises the potential energy directly through a network-parameterised displacement field rather than treating stress as an internal variable [47–50], which offers an alternative route to the same admissibility questions should the discretisation of Eqs. (32)–(34) itself be replaced by a network ansatz. Applied here, the equilibrium-based procedure gives a way to check, rather than assume, that $\hat{\mathbf{m}}$ is the material's actual symmetry axis, and to check the invariant-to-mechanism assignment of Eq. (10) itself.

A learned $\Psi_\alpha$ trained this way enters Eqs. (11)–(35) exactly as a closed-form one would: $\mathbf{S}$ and $\mathbb{C}$ follow by the same automatic differentiation used in training, the plane-stress condensation of Appendix A.5 is unaffected, and $\mathbf{K} = \mathbf{K}_{mat} + \mathbf{K}_{geo}$ still requires both terms. The geometric term has no counterpart in the formulations cited above, which carry no director field, but it does not disappear here because $\Psi_\alpha$ is learned.

### A.10 Distillation capacity of a learned representation

This study examines what happens when a learned representation is seated in the socket of Fig. 1. A corpus of $65{,}000$ constitutive states was collected from eleven solves, each state admitted only if it passed Tier A and Tier B at the point where it was recorded, which is the Tier C criterion of Section 6 applied literally. Fig. A.1(a) shows the coverage in the $(I_t, J_t)$ plane of the inter-fibre channel, with the training support and a held-out solve drawn over it. None of the $6000$ held-out states falls outside the training box, so the held-out error measures interpolation rather than extrapolation. Reporting the support explicitly is what distinguishes a held-out error that means something from one that measures the same region twice.

Panel (b) gives the relative error of the distilled network per channel: $0.013\%$ on the training set and $0.165\%$ on the held-out solve for the fibre channel, and $0.337\%$ and $0.605\%$ for the inter-fibre channel. The inter-fibre channel is the harder of the two, as expected, since it depends on two invariants rather than one and its state space is strongly non-uniform, with most samples concentrated near $I_t = 0$ and a thin tail reaching the corners of the support.

**Remark 10.** Two statements must be kept apart. The first concerns capacity: an admissible container of the form required by Definition A.1, built from convex activations on the $(J_t, J_{ft})$ slots and unconstrained weights on the linear slots, reproduces the closed form to better than one per cent at interpolation scale while remaining admissible, since none of the conditions (A1)–(A6) is relaxed to obtain it. The second concerns deployment: the operator that produced every other result in this work runs with the network identically zero, so its error against the closed form is exactly zero by construction. For that reason no overlay of a learned response against a closed-form response is presented, since at a network of zero the two coincide by construction and the comparison could not fail. Predictive quality on states outside the corpus is neither established nor claimed.

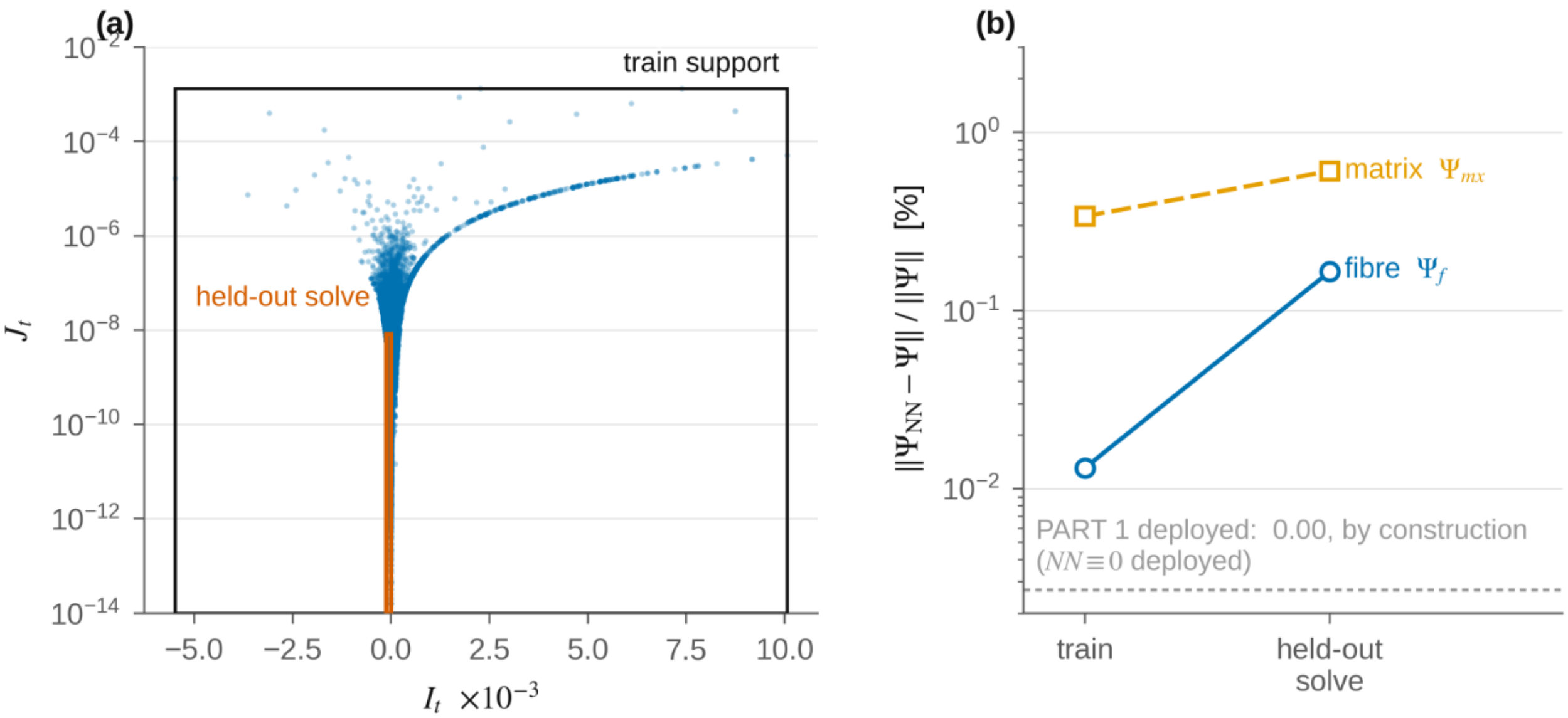

Fig. A.1. Distillation of the closed form into an admissible learned container. (a) Coverage of the 65,000 corpus states in the $(I_t, J_t)$ plane of the inter-fibre channel, with the training support and the held-out solve. (b) Relative error of the distilled network per channel on the training set and on the held-out solve; the dashed line marks the deployed configuration, whose error is zero by construction.

## Data availability

Data will be made available on request.